\documentclass[pre,aps,preprint,showpacs,preprintnumbers,amsmath,amssymb,secnumroman,eqsecnum,eqappnum]{revtex4}

\usepackage{graphicx}
\usepackage{dcolumn}
\usepackage{bm}

\newcommand{\beq}{\begin{equation}}
\newcommand{\eeq}{\end{equation}}
\newcommand{\beqa}{\begin{eqnarray}}
\newcommand{\eeqa}{\end{eqnarray}}
\newcommand{\vc}[1]{\mbox{\boldmath $#1$}}

\newcommand{\vol}[1]{{\bf #1}}

\newcommand{\du}[1]{{\bf\sf #1}}

\begin{document}

\title{Retarded hydrodynamic interaction between two spheres immersed in a viscous incompressible fluid}

\author{B. U. Felderhof}

 \email{ufelder@physik.rwth-aachen.de}
\affiliation{Institut f\"ur Theorie der Statistischen Physik\\ RWTH Aachen University\\
Templergraben 55\\52056 Aachen\\ Germany\\
}%



\date{\today}

\begin{abstract}
Retarded or frequency-dependent hydrodynamic interactions are relevant for velocity relaxation of colloidal particles immersed in a fluid, sufficiently close that their flow patterns interfere. The interactions are also important for periodic motions, such as occur in swimming. Analytic expressions are derived for the set of scalar mobility functions of a pair of spheres. Mutual hydrodynamic interactions are evaluated in one-propagator approximation, characterized by a single Green function acting between the two spheres. Self-mobility functions are evaluated in a two-propagator approximation, characterized by a single reflection between the two spheres. The approximations should yield accurate results for intermediate and long distances between the spheres. Both translations and rotations are considered. For motions perpendicular to the line of centers there is translation-rotation coupling. Extensive use is made of Fax\'en theorems which yield the hydrodynamic force and torque acting on a sphere in an incident oscillating flow.
\end{abstract}

\maketitle
\section{\label{I}Introduction}
In many problems of physics and chemistry the dynamics of particles suspended in a fluid is influenced by hydrodynamic interactions. In principle these can be evaluated on the basis of the Navier-Stokes equations for the fluid flow. Often the fluid can be assumed to be incompressible.

On the slow time-scale of diffusion inertia of the fluid can be neglected. In this limit the dynamics of the fluid can be derived from the Stokes equations of low Reynolds number hydrodynamics \cite{1}, and the hydrodynamic interactions are embodied by a friction matrix dependent on the instantaneous configuration of particles. Equivalently one can consider the mobility matrix, which is the inverse of the friction matrix.

In particular, for two spheres with no-slip boundary conditions Batchelor \cite{2} derived the first few terms in an expansion of the mobility matrix in powers of the inverse distance between centers. In early work we derived some more terms of the expansion for mixed slip-stick boundary conditions \cite{3}, and considered also permeable spheres and rotational motions \cite{4}. Schmitz and Felderhof extended the expansion further in algebraic form \cite{5}. Later the expansion was expressed in a recursive scheme which allowed accurate numerical calculation of the pair hydrodynamic interactions in the Stokes limit \cite{6}-\cite{8}. At present numerical schemes exist which allow calculation of the steady state hydrodynamic interactions between many particles \cite{9},\cite{10}.

On the time-scale of velocity relaxation and Brownian motion one must go beyond the steady state calculations, and must consider retarded or frequency-dependent hydrodynamic interactions. The Stokes limit corresponds to zero frequency. In the following we consider oscillatory translations and rotations of two spheres. Velocity relaxation is summarized by a $12\times 12$ frequency-dependent admittance matrix, which can be expressed in terms of a mass matrix and a friction matrix. We aim at an approximate analytic calculation of the set of scalar mobility functions, the elements of the mobility matrix, as functions of distance and frequency.

It follows from the reciprocal  theorem that the admittance matrix, the friction matrix, and the mobility matrix are all symmetrical  \cite{8}. The symmetry leads to a set of reciprocity relations between the scalar mobility functions. Each of the matrices can be expressed via a multiple scattering expansion as a sum of contributions describing successive scattering between the two spheres \cite{11}. This allows decomposition of each matrix as a sum of $n$-propagator contributions, where $n$ counts the number of Green functions acting between the two spheres. It follows from the symmetry of the single sphere friction operator \cite{12} and the symmetry of the Green function that the individual $n$-propagator contributions are also symmetric.

Our approximation to the mobility matrix consists of the one-propagator contribution, which is calculated exactly, and a low order multipole approximation to the two-propagator contribution. In both calculations use is made of Fax\'en type theorems for the hydrodynamic force and torque exerted on a single sphere by an incident oscillatory flow \cite{8},\cite{13},\cite{14}.

The one-propagator contribution to the translational-translational $(tt)$ part of the mobility matrix is calculated from the primary flow of the first sphere at given frequency, as found by Stokes \cite{15}, and from the corresponding force exerted on the second sphere, as found from the Fax\'en type theorem. We show explicitly that this contribution satisfies the reciprocity relation. In an earlier calculation by Ardekani and Rangel \cite{16} for two equal spheres an approximate form of the Fax\'en type theorem, due to Maxey and Riley \cite{17}, was used, so that their result is not exact. If their calculation is extended to unequal spheres the reciprocity relation is not satisfied. The same criticism applies to the calculation of Jung and Schmid of the $tt$ hydrodynamic interaction in a compressible fluid \cite{18}.

We consider both translations and rotations of the two spheres. We calculate also the one-propagator contribution to the rotational-translational $(rt)$ part, the translational-rotational $(tr)$ part, and the rotational-rotational $(rr)$ part of the one-propagator mobility matrix. For the latter two contributions the primary flow is that of a sphere performing rotatory oscillations. All these contributions  properly satisfy the reciprocity relations.

The calculation of the two-propagator contribution to the self-mobility functions is technically demanding. It involves hydrodynamic scattering from a moving sphere \cite{12}. The expressions for the self-mobility functions are correspondingly more complicated than those for the mutual functions. Since the calculation is limited to low multipole orders of the secondary flow, the reciprocity relation between the $rt$ and $tr$ functions is satisfied only approximately.

The calculated mobility functions should be accurate at all frequencies and at intermediate and long distance between centers. At short distance higher order terms in the multiple scattering expansion must be taken into account. The corresponding two-center multipole expansion can be formulated on the basis of exact addition theorems \cite{19}, but the algorithm is quite elaborate and requires numerical evaluation \cite{20},\cite{21}. Clercx \cite{21} provided plots of the scalar mobilities for some selected short distances as functions of frequency. Hermanns formulated the multiple scattering expansion for the admittance matrix \cite{22}, and investigated in particular the behavior of the velocity relaxation functions at long time \cite{23}.

In Sec. II of this article we discuss the linear dynamics of two unequal spheres immersed in a viscous incompressible fluid. In Sec. III we consider earlier approximations to the $tt$ part of the mobility matrix. In Sec. IV we derive the analytical expressions for the mutual mobility functions in one-propagator approximation. In Secs. V, VI and VII we derive the analytical expressions for the self-mobility functions in two-propagator approximation. In Sec. VIII we discuss results for both mutual and self-mobility functions. In Sec. IX we calculate the mutual functions also for mixed slip-stick boundary conditions. We finish the article with conclusions in Sec. X.

\section{\label{II}Linear response of two spheres}

We consider two spheres labeled $A,B$ of radii $a$ and $b$ centered at positions $\vc{R}_A$ and $\vc{R}_B$ and immersed in a viscous incompressible fluid of shear viscosity $\eta$ and mass density $\rho$.  The spheres are subjected to applied forces $\vc{E}_A(t),\;\vc{E}_B(t)$ and torques $\vc{N}_A(t),\;\vc{N}_B(t)$ which oscillate at frequency $\omega$. At the surface of each sphere the fluid flow velocity satisfies the no-slip boundary condition. The resulting flow velocity $\vc{u}(\vc{r},t)$ and pressure $p(\vc{r},t)$ of the fluid are assumed to satisfy the Navier-Stokes equations
\begin{equation}
\label{2.1}\rho\bigg[\frac{\partial\vc{u}}{\partial t}+\vc{u}\cdot\nabla\vc{u}\bigg]=\eta\nabla^2\vc{u}-\nabla p,\qquad\nabla\cdot\vc{u}=0.
\end{equation}
The flow velocity tends to zero at infinity and the pressure tends to the ambient value $p_0$. The equations of motion for the two spheres read
\begin{eqnarray}
\label{2.2}m_A\frac{d\vc{U}_A}{dt}&=&\vc{E}_A+\vc{K}_A,\qquad m_B\frac{d\vc{U}_B}{dt}=\vc{E}_B+\vc{K}_B,\nonumber\\
I_A\frac{d\vc{\Omega}_A}{dt}&=&\vc{N}_A+\vc{T}_A,\qquad I_B\frac{d\vc{\Omega}_B}{dt}=\vc{N}_B+\vc{T}_B,
\end{eqnarray}
where $m_A,m_B$ are the masses, $\vc{U}_A,\vc{U}_B$ are the translational velocities, and $\vc{K}_A,\vc{K}_B$ are the forces exerted by the fluid on the two spheres. Furthermore $I_A,I_B$ are the moments of inertia, $\vc{\Omega}_A,\vc{\Omega}_B$ are the rotational velocities, and $\vc{T}_A,\vc{T}_B$ are the torques exerted by the fluid on the two spheres. We assume the spheres to be uniform with mass densities $\rho_A$ and $\rho_B$, so that $m_A=4\pi\rho_Aa^3/3,\;m_B=4\pi\rho_Bb^3/3$ and $I_A=\frac{2}{5}m_A a^2,\;I_B=\frac{2}{5}m_B b^2$.

For small amplitude of the applied forces and torques the spheres perform small oscillations about rest positions $\vc{R}_A$ and $\vc{R}_B$. The relative position vector is denoted as $\vc{R}=\vc{R}_B-\vc{R}_A$. One may linearize the Navier-Stokes equations by omitting the inertial term $\rho\vc{u}\cdot\nabla\vc{u}$ and calculate the hydrodynamic forces $\vc{K}_A,\vc{K}_B$ and torques $\vc{T}_A,\vc{T}_B$ from the linearized equations. In principle the calculation can be performed exactly in terms of a multiple scattering expansion, and leads to a linear relation between sphere translational and rotational velocities and applied forces and torques. In complex notation the relation reads
\begin{eqnarray}
\label{2.3}\vc{U}_{A\omega}&=&\vc{Y}^{tt}_{AA}(\vc{R},\omega)\cdot\vc{E}_{A\omega}+\vc{Y}^{tt}_{AB}(\vc{R},\omega)\cdot\vc{E}_{B\omega}
+\vc{Y}^{tr}_{AA}(\vc{R},\omega)\cdot\vc{N}_{A\omega}+\vc{Y}^{tr}_{AB}(\vc{R},\omega)\cdot\vc{N}_{B\omega},\nonumber\\
\vc{U}_{B\omega}&=&\vc{Y}^{tt}_{BA}(\vc{R},\omega)\cdot\vc{E}_{A\omega}+\vc{Y}^{tt}_{BB}(\vc{R},\omega)\cdot\vc{E}_{B\omega}
+\vc{Y}^{tr}_{BA}(\vc{R},\omega)\cdot\vc{N}_{A\omega}+\vc{Y}^{tr}_{BB}(\vc{R},\omega)\cdot\vc{N}_{B\omega},\nonumber\\
\vc{\Omega}_{A\omega}&=&\vc{Y}^{rt}_{AA}(\vc{R},\omega)\cdot\vc{E}_{A\omega}+\vc{Y}^{rt}_{AB}(\vc{R},\omega)\cdot\vc{E}_{B\omega}
+\vc{Y}^{rr}_{AA}(\vc{R},\omega)\cdot\vc{N}_{A\omega}+\vc{Y}^{rr}_{AB}(\vc{R},\omega)\cdot\vc{N}_{B\omega},\nonumber\\
\vc{\Omega}_{B\omega}&=&\vc{Y}^{rt}_{BA}(\vc{R},\omega)\cdot\vc{E}_{A\omega}+\vc{Y}^{rt}_{BB}(\vc{R},\omega)\cdot\vc{E}_{B\omega}
+\vc{Y}^{rr}_{BA}(\vc{R},\omega)\cdot\vc{N}_{A\omega}+\vc{Y}^{rr}_{BB}(\vc{R},\omega)\cdot\vc{N}_{B\omega}.\nonumber\\
\end{eqnarray}
 According to linear response theory of statistical physics the complete $12\times 12$ admittance matrix
\begin{equation}
\label{2.4}\du{Y}(\vc{R},\omega)=\left(\begin{array}{cccc}\vc{Y}^{tt}_{AA}(\vc{R},\omega)&\vc{Y}^{tr}_{AA}(\vc{R},\omega)&
\vc{Y}^{tt}_{AB}(\vc{R},\omega)&\vc{Y}^{tr}_{AB}(\vc{R},\omega)\\
\vc{Y}^{rt}_{AA}(\vc{R},\omega)&\vc{Y}^{rr}_{AA}(\vc{R},\omega)&
\vc{Y}^{rt}_{AB}(\vc{R},\omega)&\vc{Y}^{rr}_{AB}(\vc{R},\omega)\\
\vc{Y}^{tt}_{BA}(\vc{R},\omega)&\vc{Y}^{tr}_{BA}(\vc{R},\omega)&
\vc{Y}^{tt}_{BB}(\vc{R},\omega)&\vc{Y}^{tr}_{BB}(\vc{R},\omega)\\
\vc{Y}^{rt}_{BA}(\vc{R},\omega)&\vc{Y}^{rr}_{BA}(\vc{R},\omega)&
\vc{Y}^{rt}_{BB}(\vc{R},\omega)&\vc{Y}^{rr}_{BB}(\vc{R},\omega)\\
\end{array}\right)
\end{equation}
is symmetric \cite{24}.

In the theory of Brownian motion the $12\times 12$ velocity correlation matrix $\du{C}(\vc{R},t)$ is related to the admittance matrix $\du{Y}(\vc{R},\omega)$ by the fluctuation-dissipation theorem \cite{25}. With one-sided Fourier transform
\begin{equation}
\label{2.5}\hat{\du{C}}(\vc{R},\omega)=\int^\infty_0e^{i\omega t}\du{C}(\vc{R},t)\;dt
\end{equation}
this reads
\begin{equation}
\label{2.6}\hat{\du{C}}(\vc{R},\omega)=k_BT\du{Y}(\vc{R},\omega),
\end{equation}
with Boltzmann's constant $k_B$ and absolute temperature $T$.

In the linear theory the equations of motion Eq. (2.2) become
\begin{eqnarray}
\label{2.7}-i\omega m_A\vc{U}_{A\omega}&=&\vc{E}_{A\omega}+\vc{K}_{A\omega},\qquad -i\omega m_B\vc{U}_{B\omega}=\vc{E}_{B\omega}+\vc{K}_{B\omega},\nonumber\\
-i\omega I_A\vc{\Omega}_{A\omega}&=&\vc{N}_{A\omega}+\vc{T}_{A\omega},\qquad -i\omega I_B\vc{\Omega}_{B\omega}=\vc{N}_{B\omega}+\vc{T}_{B\omega},
\end{eqnarray}
with hydrodynamic forces and torques related to the velocities by friction tensors as
\begin{eqnarray}
\label{2.8}\vc{K}_{A\omega}&=&-\vc{\zeta}^{tt}_{AA}(\vc{R},\omega)\cdot\vc{U}_{A\omega}-\vc{\zeta}^{tt}_{AB}(\vc{R},\omega)\cdot\vc{U}_{B\omega}
-\vc{\zeta}^{tr}_{AA}(\vc{R},\omega)\cdot\vc{\Omega}_{A\omega}-\vc{\zeta}^{tr}_{AB}(\vc{R},\omega)\cdot\vc{\Omega}_{B\omega},\nonumber\\
\vc{K}_{B\omega}&=&-\vc{\zeta}^{tt}_{BA}(\vc{R},\omega)\cdot\vc{U}_{A\omega}-\vc{\zeta}^{tt}_{BB}(\vc{R},\omega)\cdot\vc{U}_{B\omega}
-\vc{\zeta}^{tr}_{BA}(\vc{R},\omega)\cdot\vc{\Omega}_{A\omega}-\vc{\zeta}^{tr}_{BB}(\vc{R},\omega)\cdot\vc{\Omega}_{B\omega},\nonumber\\
\vc{T}_{A\omega}&=&-\vc{\zeta}^{rt}_{AA}(\vc{R},\omega)\cdot\vc{U}_{A\omega}-\vc{\zeta}^{rt}_{AB}(\vc{R},\omega)\cdot\vc{U}_{B\omega}
-\vc{\zeta}^{rr}_{AA}(\vc{R},\omega)\cdot\vc{\Omega}_{A\omega}-\vc{\zeta}^{rr}_{AB}(\vc{R},\omega)\cdot\vc{\Omega}_{B\omega},\nonumber\\
\vc{T}_{B\omega}&=&-\vc{\zeta}^{rt}_{BA}(\vc{R},\omega)\cdot\vc{U}_{A\omega}-\vc{\zeta}^{rt}_{BB}(\vc{R},\omega)\cdot\vc{U}_{B\omega}
-\vc{\zeta}^{rr}_{BA}(\vc{R},\omega)\cdot\vc{\Omega}_{A\omega}-\vc{\zeta}^{rr}_{BB}(\vc{R},\omega)\cdot\vc{\Omega}_{B\omega}.\nonumber\\
\end{eqnarray}
The $12\times 12$ friction matrix defined by these linear relations is symmetric. The symmetry may be derived as a consequence of the reciprocal theorem of linear hydrodynamics \cite{4},\cite{8}. The symmetry also follows from the multiple scattering expansion combined with the symmetry of the two single sphere friction operators \cite{12} and the symmetry of the Green function. We note that the friction matrix follows from the solution of the linearized Navier-Stokes equations for specified translational and rotational velocities of the two spheres under the condition that the flow disturbance tends to zero at infinity. The friction matrix is independent of the mass density of the two spheres.

The translational friction tensors defined in Eq. (2.8) include high frequency behavior proportional to $-i\omega$. Subtracting this we obtain
\begin{equation}
\label{2.9}\vc{\zeta}^{tt}_{ij}(\vc{R},\omega)=-i\omega\vc{m}_{aij}(\vc{R})+\vc{\zeta}^{tt\prime}_{ij}(\vc{R},\omega),
\qquad (i,j)=(A,B),
\end{equation}
with added mass tensors $\vc{m}_{aij}(\vc{R})$ and remaining friction tensors $\vc{\zeta}^{tt\prime}_{ij}(\vc{R},\omega)$.

We write Eq. (2.8) in abbreviated form as
\begin{equation}
\label{2.10}\left(\begin{array}{c}\du{K}
\\\du{T}
\end{array}\right)=-\left(\begin{array}{cc}\vc{\zeta}^{tt}&\vc{\zeta}^{tr}
\\\vc{\zeta}^{rt}&\vc{\zeta}^{rr}
\end{array}\right)\cdot\left(\begin{array}{c}\du{U}
\\\du{\Omega}
\end{array}\right),
\end{equation}
with $6\times 6$ matrices $\vc{\zeta}^{mn}$. The inverse relation
\begin{equation}
\label{2.11}\left(\begin{array}{c}\du{U}
\\\du{\Omega}
\end{array}\right)=-\left(\begin{array}{cc}\vc{\mu}^{tt}&\vc{\mu}^{tr}
\\\vc{\mu}^{rt}&\vc{\mu}^{rr}
\end{array}\right)\cdot\left(\begin{array}{c}\du{K}
\\\du{T}
\end{array}\right),
\end{equation}
defines the $6\times 6$ mobility matrices $\vc{\mu}^{mn}$. The $12\times 12$ mobility matrix $\vc{\mu}$ is symmetric. Cichocki and Felderhof \cite{26} derived the low frequency expansion
\begin{eqnarray}
\label{2.12}\vc{\mu}^{tt}_{ij}(\vc{R},\omega)&=&\vc{\mu}^{tt}_{ij}(\vc{R},0)-\frac{\alpha}{6\pi\eta}\vc{I}+O(\alpha^2),\qquad
\vc{\mu}^{tr}_{ij}(\vc{R},\omega)=\vc{\mu}^{tr}_{ij}(\vc{R},0)+O(\alpha^2),\nonumber\\
\vc{\mu}^{rt}_{ij}(\vc{R},\omega)&=&\vc{\mu}^{rt}_{ij}(\vc{R},0)+O(\alpha^2),\qquad
\vc{\mu}^{rr}_{ij}(\vc{R},\omega)=\vc{\mu}^{rr}_{ij}(\vc{R},0)+O(\alpha^2),\qquad (i,j)=(A,B),\nonumber\\
\end{eqnarray}
with variable $\alpha$ defined by
\begin{equation}
\label{2.13}\alpha=(-i\omega\rho/\eta)^{1/2},\qquad \mathrm{Re}\;\alpha>0,
\end{equation}
and with unit tensor $\vc{I}$.

Our strategy in the following will be to calculate an approximation to the mobility matrix by finding approximate values for the sphere velocities for simple flow situations with specified forces and torques. The calculation will be performed analytically. By inversion the calculation also yields an approximation to the friction matrix $\vc{\zeta}=\vc{\mu}^{-1}$. Hence we can find an approximation to the admittance matrix $\du{Y}=(-i\omega\du{m}_0+\vc{\zeta})^{-1}$, where $\du{m}_0$ is the mass matrix, which is diagonal with elements $m_A,m_B,I_A,I_B$.

By isotropy the various mobility tensors take the form
\begin{eqnarray}
\label{2.14}\vc{\mu}^{tt}_{ij}(\vc{R},\omega)&=&\alpha^{tt}_{ij}(R,\omega)\hat{\vc{R}}\hat{\vc{R}}+\beta^{tt}_{ij}(R,\omega)(\vc{I}-\hat{\vc{R}}\hat{\vc{R}}),\nonumber\\
\vc{\mu}^{tr}_{ij}(\vc{R},\omega)&=&\beta^{tr}_{ij}(R,\omega)\vc{\epsilon}\cdot\hat{\vc{R}},\qquad
\vc{\mu}^{rt}_{ij}(\vc{R},\omega)=\beta^{rt}_{ij}(R,\omega)\vc{\epsilon}\cdot\hat{\vc{R}},\nonumber\\
\vc{\mu}^{rr}_{ij}(\vc{R},\omega)&=&\alpha^{rr}_{ij}(R,\omega)\hat{\vc{R}}\hat{\vc{R}}+\beta^{rr}_{ij}(R,\omega)(\vc{I}-\hat{\vc{R}}\hat{\vc{R}}),\qquad (i,j)=(A,B),
\end{eqnarray}
where $\hat{\vc{R}}=\vc{R}/R$, $\vc{I}$ is the unit tensor, and $\vc{\epsilon}$ is the Levi-Civita tensor. The scalar functions satisfy the reciprocity relations
\begin{eqnarray}
\label{2.15}\alpha^{tt}_{ij}(R,\omega)&=&\alpha^{tt}_{ji}(R,\omega)\qquad\beta^{tt}_{ij}(R,\omega)=\beta^{tt}_{ji}(R,\omega),\nonumber\\
\beta^{tr}_{ij}(R,\omega)&=&-\beta^{rt}_{ji}(R,\omega),\nonumber\\
\alpha^{rr}_{ij}(R,\omega)&=&\alpha^{rr}_{ji}(R,\omega)\qquad\beta^{rr}_{ij}(R,\omega)=\beta^{rr}_{ji}(R,\omega),\qquad (i,j)=(A,B),
\end{eqnarray}
as follows from the symmetry of the $12\times 12$ mobility matrix $\vc{\mu}$.

It is convenient to use Cartesian coordinates with the origin at $\vc{R}_A$ and $z$ axis in the direction of $\hat{\vc{R}}$. In this representation it is evident that translational motions along the line of centers and rotations about this line decouple from translational motions perpendicular to the line and rotations about an axis perpendicular to the line of centers. We call these two types of motion longitudinal and transverse, and denote them by the symbols $\parallel$ and $\perp$. By permutation of rows and columns the mobility matrix can be rearranged in the decomposed form
\begin{equation}
\label{2.16}\vc{\mu}_D=\left(\begin{array}{ccc}\vc{\mu}_\perp&\vc{0}&\vc{0}\\
\vc{0}&\vc{\mu}^-_\perp&\vc{0}\\
\vc{0}&\vc{0}&\vc{\mu}_\parallel\\
\end{array}\right),
\end{equation}
where the transverse matrix
\begin{equation}
\label{2.17}\vc{\mu}_\perp=\left(\begin{array}{cccc}\beta^{tt}_{AA}&\beta^{tt}_{AB}&\beta^{tr}_{AA}&\beta^{tr}_{AB}\\
\beta^{tt}_{AB}&\beta^{tt}_{BB}&\beta^{tr}_{AB}&\beta^{tr}_{BB}\\
\beta^{tr}_{AA}&\beta^{tr}_{AB}&\beta^{rr}_{AA}&\beta^{rr}_{AB}\\
\beta^{tr}_{AB}&\beta^{tr}_{BB}&\beta^{rr}_{AB}&\beta^{rr}_{BB}
\end{array}\right),
\end{equation}
corresponds to translational motions and forces in the $x$ direction and rotational motions and torques in the $y$ direction. We have used the reciprocity relations Eq. (2.15). The second matrix in Eq. (2.16) is given by
\begin{equation}
\label{2.18}\vc{\mu}^-_\perp=\left(\begin{array}{cccc}\beta^{tt}_{AA}&\beta^{tt}_{AB}&-\beta^{tr}_{AA}&-\beta^{tr}_{AB}\\
\beta^{tt}_{AB}&\beta^{tt}_{BB}&-\beta^{tr}_{AB}&-\beta^{tr}_{BB}\\
-\beta^{tr}_{AA}&-\beta^{tr}_{AB}&\beta^{rr}_{AA}&\beta^{rr}_{AB}\\
-\beta^{tr}_{AB}&-\beta^{tr}_{BB}&\beta^{rr}_{AB}&\beta^{rr}_{BB}
\end{array}\right),
\end{equation}
and corresponds to translational motions and forces in the $y$ direction and rotational motions and torques in the $x$ direction. It can be transformed to $\vc{\mu}_\perp$ by referring rotational motions and torques to the negative $x$ direction, corresponding to axial symmetry.
The longitudinal mobility matrix
\begin{equation}
\label{2.19}\vc{\mu}_\parallel=\left(\begin{array}{cccc}\alpha^{tt}_{AA}&\alpha^{tt}_{AB}&0&0\\
\alpha^{tt}_{AB}&\alpha^{tt}_{BB}&0&0\\
0&0&\alpha^{rr}_{AA}&\alpha^{rr}_{AB}\\
0&0&\alpha^{rr}_{AB}&\alpha^{rr}_{BB}\\
\end{array}\right),
\end{equation}
corresponds to translational motions and forces and rotational motions and torques in the $z$ direction. It is evident that for longitudinal motions translations and rotations decouple. The expressions in Eqs. (2.17-19) show that only a limited number of scalar functions need be calculated.

\section{\label{III}Approximate calculations}

The mobility matrix $\vc{\mu}(\vc{R},\omega)$ or the friction matrix $\vc{\zeta}(\vc{R},\omega)$, may be calculated in terms of a multiple scattering expansion, but the calculation is complex. In the literature final results are often presented in graphical form \cite{20}-\cite{22}. The response of a single sphere to an applied force and an incident flow is described by a multipolar friction matrix \cite{12}. Propagation between spheres is described by the Green function of the linearized Navier-Stokes equations
\begin{equation}
\label{3.1}\eta[\nabla^2\vc{v}-\alpha^2\vc{v}]-\nabla p=0,\qquad\nabla\cdot\vc{v}=0,\qquad (r>0).
\end{equation}
The Green function reads
\begin{equation}
\label{3.2}\vc{G}(\vc{r},\omega)=G_{\parallel}(r,\omega)\hat{\vc{r}}\hat{\vc{r}}+G_{\perp}(r,\omega)(\vc{I}-\hat{\vc{r}}\hat{\vc{r}}),
\end{equation}
with $\hat{\vc{r}}=\vc{r}/r$ and longitudinal and transverse parts
\begin{eqnarray}
\label{3.3}G_{\parallel}(r,\omega)&=&\frac{1}{2\pi\eta\alpha^2r^3}[1-(1+\alpha r)e^{-\alpha r}],\nonumber\\
G_{\perp}(r,\omega)&=&\frac{-1}{4\pi\eta\alpha^2r^3}[1-(1+\alpha r+\alpha^2r^2)e^{-\alpha r}].
\end{eqnarray}

Van Saarloos and Mazur \cite{27} performed the multiple scattering expansion on the basis of a spatial Fourier transform and derived the first few terms in an expansion of the mobility matrix in powers of inverse distance $R^{-1}$. The $tt$ mutual interaction, presented in Eq. (4.3) of their article, does not satisfy the reciprocity relation Eq. (2.15), and is therefore not acceptable. Their expression for the self-interaction does not depend on mutual distance.

Pie\'nkowska \cite{28} derived an approximate expression for the mutual friction tensor, valid at low frequency, which satisfies reciprocity. Later she derived a more detailed expression, which again satisfies reciprocity \cite{29}. Her expression is closely related to the Green function.

Henderson et al. \cite{30} considered equal spheres and only the longitudinal part of the friction tensor, but their expression is easily generalized to the complete tensor and to unequal spheres. The mutual friction tensor is approximated as
\begin{equation}
\label{3.4}\vc{\zeta}^{tt}_{AB}(\vc{R},\omega)=\vc{\zeta}^{tt}_{BA}(\vc{R},\omega)\approx-(6\pi\eta)^2ab\vc{G}(\vc{R},\omega),
\end{equation}
and the single sphere admittances as
\begin{equation}
\label{3.5}\mathcal{Y}^t_{A\parallel}=\mathcal{Y}^t_{A\perp}\approx\frac{1}{6\pi\eta a(1+\alpha a)},
\end{equation}
and similarly for $B$. We recall that the frequency-dependent translational friction coefficient of a single sphere is given by
\begin{equation}
\label{3.6}\zeta^t_A(\omega)=6\pi\eta a\big(1+\alpha a+\frac{1}{9}\alpha^2a^2\big),
\end{equation}
where the last term corresponds to the added mass. Hence the single sphere admittance is
\begin{equation}
\label{3.7}\mathcal{Y}^t_A=[-i\omega(m_A+\frac{1}{2}m_{fA})+6\pi\eta a(1+\alpha a)]^{-1},
\end{equation}
where $m_{fA}=4\pi\rho a^3/3$ is the mass of fluid displaced by $A$. Clearly the high frequency behavior is not well described by the approximation Eq. (3.5). Henderson et al. \cite {30} find good agreement with their experiments in the low frequency regime. A better approximation would be obtained by use of Eq. (3.7) instead of Eq. (3.5).

Bonet Avalos et al. studied the role of time-dependent hydrodynamic interactions on the dynamics of polymers in solution \cite{31}. They made a point approximation in which the mutual mobility tensor is given by
\begin{equation}
\label{3.8}\vc{\mu}^{tt}_{AB}(\vc{R},\omega)=\vc{\mu}^{tt}_{BA}(\vc{R},\omega)\approx\vc{G}(\vc{R},\omega).
\end{equation}
The approximation satisfies reciprocity, but is clearly rather drastic. Ardekani and Rangel \cite{16} in their first approach, and Tatsumi and Yamamoto \cite{32} made the same approximation. The approximation was recommended for use in two-point passive microrheology by C\'ordoba et al. \cite{33}. In their second approach Ardekani and Rangel \cite{16} attempted to derive a finite-size correction for the translational friction functions, but they did not take proper account of the expression for the Fax\'en type expression for the force on a sphere in a non-uniform flow, as discussed below.

The calculations below are based on an approximation in which the mutual terms involve only a single propagator between spheres and the self terms of the mobility matrix are approximated by their single sphere expressions corrected by at most one reflection from the other sphere. Finite-size corrections are taken into account. Later we shall compare with the point approximation Eq. (3.8).

\section{\label{IV}Mutual mobility functions}

 Our calculation is based on the method of reflections and amounts to an approximate solution of the linearized Navier-Stokes equations as a boundary value problem. We employed the same method earlier to calculate an approximate mobility matrix at zero frequency \cite{3}. We start from a given set of translational and rotational velocities $\vc{U}_{A,0},\vc{U}_{B,0},\vc{\Omega}_{A,0},\vc{\Omega}_{B,0}$. To lowest order this corresponds to a set of hydrodynamic forces $\vc{K}_{A,0},\vc{K}_{B,0}$ and torques $\vc{T}_{A,0},\vc{T}_{B,0}$, as given by the single sphere translational and rotational friction coefficients
 \begin{eqnarray}
\label{4.1}
\zeta^t_A(\omega)&=&6\pi\eta aA_0(\alpha a),\qquad\zeta^t_B(\omega)=6\pi\eta bA_0(\alpha b),\nonumber\\
\zeta^r_A(\omega)&=&8\pi\eta a^3\frac{B_0(\alpha a)}{1+\alpha a},\qquad\zeta^r_B(\omega)=8\pi\eta b^3\frac{B_0(\alpha b)}{1+\alpha b},
\end{eqnarray}
with abbreviations $A_0(\lambda)$ and $B_0(\lambda)$ given in the Appendix.
Higher order corrections $\vc{U}_{A,j},\vc{U}_{B,j},\vc{\Omega}_{A,j},\vc{\Omega}_{B,j}$ $(j=1,2,...)$ are calculated from the condition that in each step of the calculation the corresponding forces and torques vanish. This is achieved by the application of Fax\'en type theorems.

Fax\'en's theorem for the hydrodynamic force acting on a sphere of radius $a$ with center at the origin and subject to an incident flow $\vc{v}_0(\vc{r},\omega),p_0(\vc{r},\omega)$, which is a solution of the linearized Navier-Stokes equations, reads \cite{8}
\begin{equation}
\label{4.2}\vc{K}_A=6\pi\eta a\big[B_0(\alpha a)\vc{v}_0+B_2(\alpha a)a^2\nabla^2\vc{v}_0\big]\big|_{\vc{r}=\vc{0}}-\zeta^t_A(\omega)\vc{U}_A,
\end{equation}
with the further abbreviation $B_2(\lambda)$ given in the Appendix. In the original formulation of the Fax\'en theorem by Mazur and Bedeaux \cite{34} the force exerted by the fluid on a single sphere is expressed in terms of surface and volume averages of the incident flow field. This makes clear that only $1mN$ and $1mP$ multipole components of the incident flow, in the notation of Felderhof and Jones \cite{12}, contribute to the force.

The analogous theorem for the hydrodynamic torque reads \cite{13},\cite{14}
\begin{equation}
\label{4.3}\vc{T}_A=4\pi\eta a^3\frac{e^{\alpha a}}{1+\alpha a}(\nabla\times\vc{v}_0)\big|_{\vc{r}=\vc{0}}-\zeta^r_A(\omega)\vc{\Omega}_A.
\end{equation}
The equivalent form which we derived \cite{35},\cite{36} is more complicated, but makes clear that only the three $1mM$ multipole components of the incident flow contribute to the torque.

Consider first the zero order situation where only $\vc{U}_{A,0}=U_{A,0}\vc{e}_z$ and $\vc{K}_{A,0}$ differ from zero. The corresponding zero order flow pattern is \cite{15},\cite{12}
\begin{equation}
\label{4.4}\vc{v}_{UA\omega,0}(\vc{r})=U_{A,0}\bigg[\frac{3B_0(\alpha a)}{2\alpha^2a^2}\;\vc{u}_{1A}(\vc{r})+\frac{1}{2}\;\alpha a\;\vc{v}_{1A}(\vc{r},\alpha)\bigg],
\end{equation}
where $\vc{u}_{1A}(\vc{r})$ is a potential flow and $\vc{v}_{1A}(\vc{r},\alpha)$ is a viscous flow. Explicitly
 \begin{eqnarray}
\label{4.5}
\vc{u}_{1A}(\vc{r})&=&-\bigg(\frac{a}{r}\bigg)^{3}\vc{B}_{1z}(\hat{\vc{r}}),\nonumber\\
\vc{v}_{1A}(\vc{r},\alpha)&=&\frac{2}{\pi}\;e^{\alpha a}[2k_{0}(\alpha r)\vc{A}_{1z}(\hat{\vc{r}})+k_{2}(\alpha r)\vc{B}_{1z}(\hat{\vc{r}})],
\end{eqnarray}
with modified spherical Bessel functions \cite{37} $k_l(z)$, and vector spherical harmonics given by
\begin{equation}
\label{4.6}\vc{A}_{1z}(\hat{\vc{r}})=\vc{e}_z,\qquad\vc{B}_{1z}(\hat{\vc{r}})=\vc{e}_z-3(\vc{e}_z\cdot\hat{\vc{r}})\hat{\vc{r}}.
\end{equation}
The flows $\vc{u}_{1A}$ and $\vc{v}_{1A}$ satisfy the equations
\begin{eqnarray}
\label{4.7}\nabla^2\vc{u}_{1A}&=&0,\qquad\nabla\cdot\vc{u}_{1A}=0,\nonumber\\
\nabla^2\vc{v}_{1A}-\alpha^2\vc{v}_{1A}&=&0,\qquad\nabla\cdot\vc{v}_{1A}=0,\qquad (r>0).
\end{eqnarray}
In the notation of Felderhof and Jones \cite{12}
\begin{equation}
\label{4.8}\vc{u}_{1A}(\vc{r})=2\sqrt{3\pi}\;\alpha^2a^3\vc{v}^-_{10P}(\vc{r}),\qquad
\vc{v}_{1A}(\vc{r})=4\sqrt{3\pi}\;\frac{e^{\alpha a}}{\alpha}\vc{v}^-_{10N}(\vc{r}),
\end{equation}
with functions
 \begin{eqnarray}
\label{4.9}
\vc{v}^-_{lmN}(\vc{r})&=&\frac{2\alpha}{\pi l(l+1)(2l+1)}\;[(l+1)k_{l-1}(\alpha r)\vc{A}_{lm}(\hat{\vc{r}})+k_{l+1}(\alpha r)\vc{B}_{lm}(\hat{\vc{r}})],\nonumber\\
\vc{v}^-_{lmP}(\vc{r})&=&\frac{-1}{\alpha^2(2l+1)}\;r^{-l-2}\vc{B}_{lm}(\hat{\vc{r}}),
\end{eqnarray}
with vector spherical harmonic $\vc{A}_{lm}(\hat{\vc{r}})$ and $\vc{B}_{lm}(\hat{\vc{r}})$.
It is checked by use of Eq. (4.8) that the flow pattern in Eq. (4.4) satisfies the no-slip boundary condition
\begin{equation}
\label{4.10}\vc{v}_{UA\omega,0}(\vc{r})\big|_{r=a}=U_{A,0}\vc{e}_z.
\end{equation}

To first order sphere $B$ moves with a velocity $\vc{U}_{B,1}=U_{B,1}\vc{e}_z$ such that it exerts no force on the fluid. By use of the Fax\'en theorem Eq. (4.2)
\begin{equation}
\label{4.11}U_{B,1}=\zeta^t_A(\omega)\alpha^{tt}_{BA,1}(R,\omega)U_{A,0},
\end{equation}
with
\begin{equation}
\label{4.12}\alpha^{tt}_{BA,1}(R,\omega)=\frac{B_0(\alpha a)B_0(\alpha b)-(1+\alpha R)e^{\alpha(a+b-R)}}{2\pi\eta\alpha^2R^3A_0(\alpha a)A_0(\alpha b)}.
\end{equation}
Since $K_{A,0}=-\zeta^t_A(\omega)U_{A,0}$, this may be identified with the first order scalar mobility function. The reciprocity relation $\alpha^{tt}_{BA}(R,\omega)=\alpha^{tt}_{AB}(R,\omega)$ is obviously satisfied in the approximation. At low frequency Eq. (4.12) becomes
\begin{equation}
\label{4.13}\alpha^{tt}_{AB}(R,\omega)=\alpha^{tt}_{BA}(R,\omega)=\frac{1}{4\pi\eta R}-\frac{a^2+b^2}{12\pi\eta R^3}-\frac{\alpha}{6\pi\eta}+O(\alpha^2).
\end{equation}
The zero frequency limit agrees to terms of order $R^{-6}$ with earlier results \cite{3}\cite{4}. By use of the Fax\'en theorem Eq. (4.3) one finds that in this situation the rotational velocity of sphere $B$ vanishes, so that $\alpha^{rt}_{BA,1}(R,\omega)=0$.

By the same derivation for motion in the $x$ direction one finds the transverse mobility function
\begin{equation}
\label{4.14}\beta^{tt}_{BA,1}(R,\omega)=\frac{-B_0(\alpha a)B_0(\alpha b)+A_1(\alpha R)e^{\alpha(a+b-R)}}{4\pi\eta\alpha^2R^3A_0(\alpha a)A_0(\alpha b)},
\end{equation}
with abbreviation $A_1(\lambda)$ given in the Appendix. At low frequency this becomes
\begin{equation}
\label{4.15}\beta^{tt}_{AB}(R,\omega)=\beta^{tt}_{BA}(R,\omega)=\frac{1}{8\pi\eta R}+\frac{a^2+b^2}{24\pi\eta R^3}-\frac{\alpha}{6\pi\eta}+O(\alpha^2).
\end{equation}
The zero frequency limit agrees to terms of order $R^{-10}$ with earlier results \cite{3}-\cite{5}. One sees also that Eq. (3.8) is an approximation to Eqs. (4.12) and (4.14) valid for $\alpha a<<1$ and $\alpha b<<1$.

Calculating the rotational velocity of sphere $B$ about the $y$ axis one finds in this situation by use of the Fax\'en theorem Eq. (4.3) the mobility function
\begin{equation}
\label{4.16}
\beta^{rt}_{BA}(R,\omega)=\frac{1}{8\pi\eta R^2A_0(\alpha a)B_0(\alpha b)}\;(1+\alpha R)e^{\alpha(a+b-R)}.
\end{equation}
At low frequency this becomes
\begin{equation}
\label{4.17}\beta^{rt}_{BA}(R,\omega)=\frac{1}{8\pi\eta R^2}+O(\alpha^2).
\end{equation}
The zero frequency limit agrees to terms of order $R^{-9}$ with earlier results \cite{4},\cite{5}.

Next we consider a situation where in zero order sphere $A$ rotates about the $z$ axis with angular velocity $\Omega_{A,0}$ corresponding to hydrodynamic torque $T_{A,0}$. The flow pattern about sphere $A$ for rotation in the $z$ direction in the absence of incident flow is given by
\begin{equation}
\label{4.18}\vc{v}_{\Omega A\omega,0}(\vc{r})=\Omega_{A,0} a\frac{k_1(\alpha r)}{k_1(\alpha a)}\vc{C}_{1z}(\hat{\vc{r}}),
\end{equation}
with vector spherical harmonic $\vc{C}_{1z}(\hat{\vc{r}})=\vc{e}_z\times\vc{e}_r$.
The flow pattern satisfies the no-slip boundary condition
\begin{equation}
\label{4.19}\vc{v}_{\Omega A\omega,0}(\vc{r})\big|_{r=a}=\Omega_{A,0} a\vc{e}_z\times\vc{e}_r,
\end{equation}
and the equations
\begin{equation}
\label{4.20}\nabla^2\vc{v}_{\Omega A\omega,0}-\alpha^2\vc{v}_{\Omega A\omega,0}=0,\qquad\nabla\cdot\vc{v}_{\Omega A\omega,0}=0.
\end{equation}
Calculating the resulting first order rotational velocity of sphere $B$ by use of the Fax\'en theorem Eq. (4.3) we find the scalar mobility function
\begin{equation}
\label{4.21}\alpha^{rr}_{BA,1}(R,\omega)=\frac{(1+\alpha R)e^{\alpha(a+b-R)}}{8\pi\eta R^3B_0(\alpha a)B_0(\alpha b)}.
\end{equation}
This clearly satisfies reciprocity $\alpha^{rr}_{AB}(R,\omega)=\alpha^{rr}_{BA}(R,\omega)$. At low frequency
\begin{equation}
\label{4.22}\alpha^{rr}_{AB}(R,\omega)=\frac{1}{8\pi\eta R^3}+O(\alpha^2).
\end{equation}
The zero frequency limit agrees to terms of order $R^{-12}$ with earlier results \cite{4},\cite{5}. By use of the Fax\'en theorem Eq. (4.2) one finds that in this situation the translational velocity of sphere $B$ vanishes, so that $\alpha^{tr}_{BA,1}(R,\omega)=0$.

By the same derivation for rotation about the $x$ direction one finds the transverse mobility function
\begin{equation}
\label{4.23}\beta^{rr}_{AB,1}(R,\omega)=\frac{-A_1(\alpha R)e^{\alpha(a+b-R)}}{16\pi\eta R^3B_0(\alpha a)B_0(\alpha b)}.
\end{equation}
At low frequency this becomes
\begin{equation}
\label{4.24}\beta^{rr}_{AB}(R,\omega)=\beta^{rr}_{BA}(R,\omega)=\frac{-1}{16\pi\eta R^3}+O(\alpha^2).
\end{equation}
The zero frequency limit agrees to terms of order $R^{-8}$ with earlier results \cite{4}\cite{5}. Calculating the function $\beta^{tr}_{BA,1}(r,\omega)$ in this situation one finds
\begin{equation}
\label{4.25}
\beta^{tr}_{BA}(R,\omega)=\frac{-1}{8\pi\eta R^2B_0(\alpha a)A_0(\alpha b)}\;(1+\alpha R)e^{\alpha(a+b-R)},
\end{equation}
quite similar to Eq. (4.16), and in accordance with the reciprocity relation Eq. (2.15).

\section{\label{V}Longitudinal self-mobility functions}

In the calculation of the self-mobility functions we carry the method of reflections one order higher. We consider first the self-mobility function $\alpha^{tt}_{AA}(R,\omega)$. In lowest approximation we calculate this from the velocity of sphere $A$ caused by the primary flow field reflected once from sphere $B$. In the theory of Felderhof and Jones \cite{12} the reflection is described as hydrodynamic scattering by the moving sphere $B$. We shall take account of multipoles of the incident flow on sphere $B$ only up to order $l=2$.

Thus we expand the zero order flow $\vc{v}_{UA\omega,0}(\vc{r})$, centered at $\vc{R}_A=\vc{0}$, in terms of multipole flows centered at $\vc{R}_B$ as
\begin{eqnarray}
\label{5.1}\vc{v}_{UA\omega,0}(\vc{r})&=&c^+_{10N}\vc{v}^+_{10N}(\vc{r}-\vc{R}_B)+c^+_{10P}\vc{v}^+_{10P}(\vc{r}-\vc{R}_B)\nonumber\\
&+&c^+_{20N}\vc{v}^+_{20N}(\vc{r}-\vc{R}_B)
+c^+_{20P}\vc{v}^+_{20P}(\vc{r}-\vc{R}_B)+...,
\end{eqnarray}
with superposition coefficients which can be calculated from the flow given by Eq. (4.4) and with functions
 \begin{eqnarray}
\label{5.2}
\vc{v}^+_{lmN}(\vc{r})&=&\frac{1}{2l+1}\big[(l+1)g_{l-1}(\alpha r)\vc{A}_{lm}(\hat{\vc{r}})+lg_{l+1}(\alpha r)\vc{B}_{lm}(\hat{\vc{r}})\big],\nonumber\\
\vc{v}^+_{lmP}(\vc{r})&=&r^{l-1}\vc{A}_{lm}(\hat{\vc{r}}).
\end{eqnarray}
As shown in Eqs. (4.4-9) the flow $\vc{v}_{UA\omega,0}(\vc{r})$ is a linear combination of outgoing waves $\vc{v}^-_{10N}(\vc{r})$ and $\vc{v}^-_{10P}(\vc{r})$. More generally, singular solutions of this type can be expanded as the addition theorems \cite{19},\cite{38}
 \begin{eqnarray}
\label{5.3}
\vc{v}^-_{l0N}(\vc{r})&=&\sum_{l'}S^{+-}(R;l'0N,l0N)\vc{v}^+_{l'0N}(\vc{r}_<),\nonumber\\
\vc{v}^-_{l0P}(\vc{r})&=&\sum_{l'}S^{+-}(R;l'0P,l0P)\vc{v}^+_{l'0P}(\vc{r}_<),
\end{eqnarray}
where $\vc{r}_<=\vc{r}-R\vc{e}_z=\vc{r}-\vc{R}_B$. The coefficient functions $S^{+-}$ can be found \cite{19},\cite{38} from the elaborate results derived by Langbein \cite{39}. The coefficients $c^+_{10N}$ and $c^+_{10P}$ in Eq. (5.1) can be expressed in terms of functions $S^{+-}(R;l0N,10N)$ and $S^{+-}(R;l0P,10P)$, respectively.

For the low orders under consideration we prefer a more pedestrian approach. We use instead the Taylor expansion of the function $\vc{v}_{UA\omega,0}(\vc{r})$ in powers of $\vc{R}-\vc{R}_B$, with each term prolonged into a regular solution of the linearized Navier-Stokes equations Eq. (3.1), as
\begin{equation}
\label{5.4}\vc{v}_{UA\omega,0}(\vc{r})=\sum^\infty_{n=0}\vc{v}^{(n)}_{UAB},
\end{equation}
where $n$ indicates the power of $\vc{r}-\vc{R}_B$ in the Taylor
expansíon. The first term is
\begin{equation}
\label{5.5}\vc{v}^{(0)}_{UAB}=c^+_{10N}\vc{v}^+_{10N}(\vc{r}-\vc{R}_B)+c^+_{10P}\vc{v}^+_{10P}(\vc{r}-\vc{R}_B).
\end{equation}
At $\vc{r}=\vc{R}_B$ we have from Eq. (4.4)
\begin{equation}
\label{5.6}\vc{v}_{UA\omega,0}(\vc{R}_B)=U^{(0)}_B\vc{e}_z,\qquad U^{(0)}_B=U_{A,0}\frac{3a}{\alpha^2R^3}\big[B_0(\alpha a)-(1+\alpha R)e^{\alpha(a-R)}\big].
\end{equation}
Comparing this with Eq. (5.5) we find the relation
\begin{equation}
\label{5.7}U^{(0)}_B=c^+_{10N}\frac{1}{\sqrt{3\pi}}+c^+_{10P}\sqrt{\frac{3}{4\pi}}.
\end{equation}
It follows from the Fax\'en theorem Eq. (4.2) that
\begin{equation}
\label{5.8}B_0(\alpha b)U^{(0)}_B\vc{e}_z+B_2(\alpha b)b^2(\nabla^2\vc{v}^{(0)}_{UAB})\bigg|_{\vc{r}=\vc{R}_B}=A_0(\alpha b)U_{B,1}\vc{e}_z,
\end{equation}
with $U_{B,1}$ given by Eq. (4.11).
Only the first term on the right in Eq. (5.5) contributes to the second term in this equation. Substituting we obtain the value of $c^+_{10N}$ as
\begin{equation}
\label{5.9}c^+_{10N}=-3\sqrt{3\pi}\;U_{A,0}\frac{a}{\alpha^2R^3}(1+\alpha R)e^{\alpha(a-R)},
\end{equation}
and from Eq. (5.7) we find
\begin{equation}
\label{5.10}c^+_{10P}=\sqrt{\frac{4\pi}{3}}\;U_{A,0}\frac{3a}{\alpha^2R^3}B_0(\alpha a).
\end{equation}

The first line in Eq. (5.1) generates the scattered wave
\begin{eqnarray}
\label{5.11}\vc{v}^-_{B1}(\vc{r})&=&c^+_{10N}\big[\chi_{1NN}\vc{v}^-_{10N}(\vc{r}-\vc{R}_B)+\chi_{1NP}\vc{v}^-_{10P}(\vc{r}-\vc{R}_B)\big]\nonumber\\
&+&\big(c^+_{10P}-\sqrt{\frac{4\pi}{3}}\;U_{B,1}\big)\big[\chi_{1NP}\vc{v}^-_{10N}(\vc{r}-\vc{R}_B)+\chi_{1PP}\vc{v}^-_{10P}(\vc{r}-\vc{R}_B)\big],
\end{eqnarray}
with scattering coefficients
\begin{equation}
\label{5.12}\chi_{1NN}=\frac{1-e^{2\alpha b}}{\alpha},\qquad\chi_{1NP}=-3be^{\alpha b},\qquad\chi_{1PP}=-\frac{3}{2}\alpha^2b^3\frac{k_2(\alpha b)}{k_0(\alpha b)},
\end{equation}
and outgoing waves of types $10N$ and $10P$ given by Eq. (4.9).
In the second line of Eq. (5.11) we took account \cite{12} of the fact that sphere $B$ is moving with velocity $U_{B,1}$.

For the coefficients in the second line of Eq. (5.1) we find
\begin{equation}
\label{5.13}c^+_{20N}=0,\qquad c^+_{20P}=9\sqrt{\frac{\pi}{5}}\;U_{A,0}\frac{a}{\alpha^2R^4}\bigg[e^{\alpha(a-R)}B_0(\alpha R)-B_0(\alpha a)\bigg].
\end{equation}
The corresponding scattered wave is
\begin{equation}
\label{5.14}\vc{v}^-_{B2}(\vc{r})=
c^+_{20P}\big[\chi_{2NP}\vc{v}^-_{20N}(\vc{r}-\vc{R}_B)+\chi_{2PP}\vc{v}^-_{20P}(\vc{r}-\vc{R}_B)\big],
\end{equation}
with scattering coefficients
\begin{equation}
\label{5.15}
\chi_{2NP}=-5\pi\frac{b}{\alpha k_1(\alpha b)},\qquad\chi_{2PP}=-\frac{10}{3}\alpha^2b^5\frac{k_3(\alpha b)}{k_1(\alpha b)},
\end{equation}
and outgoing waves of types $20N$ and $20P$, given by Eq. (4.9).

The scattered waves act back on sphere $A$ and the corresponding velocity of sphere $A$, as its moves in this flow with zero force and torque, can be evaluated with the aid of Fax\'en theorems Eqs. (4.2) and (4.3), as above. We verify from Eq. (4.3) that the rotational velocity of sphere $A$ vanishes, corresponding to $\alpha^{rt}_{AA}(R,\omega)=0$. In our approximation the longitudinal translational self-mobility function is given by
\begin{equation}
\label{5.16}
\alpha^{tt}_{AA}(R,\omega)=\frac{1}{6\pi\eta aA_0(\alpha a)}+\alpha^{tt}_{AA,1}(R,\omega)+\alpha^{tt}_{AA,2}(R,\omega),
\end{equation}
where the first term is the single sphere contribution, the second term follows from Eq. (5.11), and the third term follows from Eq. (5.14). The explicit expression for the mobility function $\alpha^{tt}_{AA,1}(R,\omega)$ reads
\begin{eqnarray}
\label{5.17}
\alpha^{tt}_{AA,1}(R,\omega)&=&\bigg[-4\alpha^3b^3B_0(\alpha a)^2B_0(\alpha b)\nonumber\\&-&9A_0(\alpha b)(1+\alpha R)^2e^{2\alpha (a-R)}
+8\alpha^3b^3B_0(\alpha a)(1+\alpha R)e^{\alpha(a+b-R)}\nonumber\\&+&(9-9\alpha b+\alpha^2 b^2)(1+\alpha R)^2e^{2\alpha(a+b-R)}\bigg]
\bigg/\bigg(12\pi\eta\alpha^5R^6A_0(\alpha a)^2A_0(\alpha b)\bigg).\nonumber\\
\end{eqnarray}
At low frequency this has the expansion
\begin{equation}
\label{5.18}
\alpha^{tt}_{AA,1}(R,\omega)=\frac{b^5}{30\pi\eta R^6}+O(\alpha^2).
\end{equation}
We note that if the term with $U_{B,1}$ in Eq. (5.11) were omitted, corresponding to a fixed sphere $B$, the function would have a contribution proportional to $R^{-2}$.

The explicit expression for the mobility function $\alpha^{tt}_{AA,2}(R,\omega)$ reads
\begin{eqnarray}
\label{5.19}
\alpha^{tt}_{AA,2}(R,\omega)&=&\frac{-3b^3}{2\pi\eta\alpha^4R^8(1+\alpha b)A_0(\alpha a)^2}\bigg(e^{\alpha(a-R)}B_0(\alpha R)-B_0(\alpha a)\bigg)\nonumber\\&\times&\bigg(15e^{\alpha(a+b-R)}B_0(\alpha R)-B_0(\alpha a)A_3(\alpha b)\bigg),
\end{eqnarray}
with abbreviation $A_3(\lambda)$ given in the Appendix. At low frequency this has the expansion
\begin{equation}
\label{5.20}\alpha^{tt}_{AA,2}(R,\omega)=\frac{1}{8\pi\eta a}\bigg[-\frac{5ab^3}{R^4}+\frac{10a^3b^3+3ab^5}{R^6}-\frac{5a^5b^3+3a^3b^5}{R^8}\bigg]
+O(\alpha^2).
\end{equation}
The zero frequency limit of the function defined in Eq. (5.16) agrees to terms of order $R^{-5}$ with earlier results \cite{4}\cite{5}, which took account of higher order terms in the equivalent of Eq. (5.1).  The term linear in $\alpha$, arising from the first term on the right in Eq. (5.16), agrees with the exact relation Eq. (2.12).

Finally we calculate an approximation to the rotational self-mobility functions, corresponding to a torque in the $z$ direction exerted on sphere $A$. The flow pattern of sphere $A$ rotating with angular velocity $\Omega_{A,0}$ about the $z$ axis is given by Eq. (4.18).

From the Fax\'en theorem Eq. (4.2), as applied to $\vc{v}_{\Omega A\omega,0}$ for the sphere centered at $\vc{R}_B$, we see that the resulting translational velocity of sphere $B$ vanishes, since $\vc{R}=R\vc{e}_z$ and $\vc{e}_z\times\vc{R}=\vc{0}$. This confirms that in the longitudinal case there is no translation-rotation coupling.

We express the Taylor expansion of the function $\vc{v}_{\Omega A\omega,0}(\vc{r})$ in powers of $\vc{R}-\vc{R}_B$, with each term prolonged into a regular solution of the linearized Navier-Stokes equations Eq. (3.1), as
\begin{equation}
\label{5.21}\vc{v}_{\Omega A\omega,0}(\vc{r})=\sum^\infty_{n=0}\vc{v}^{(n)}_{\Omega AB}.
\end{equation}
In order to find a low order approximation to the self-mobility function $\alpha^{rr}_{AA}(R,\omega)$ we calculate the rotational velocity of sphere $A$ caused by the flow field reflected from sphere $B$. The lowest order term in Eq. (5.21) vanishes, since $\vc{R}=R\vc{e}_z$ and $\vc{e}_z\times\vc{e}_z=0$. We expand the zero order flow $\vc{v}_{\Omega A\omega,0}(\vc{r})$ in terms of incident multipole flows centered at $\vc{R}_B$ as
 \begin{equation}
\label{5.22}\vc{v}_{\Omega A\omega,0}(\vc{r})=c^+_{10M}\vc{v}^+_{10M}(\vc{r}-\vc{R}_B)+c^+_{20M}\vc{v}^+_{20M}(\vc{r}-\vc{R}_B)+...,
\end{equation}
with functions
\begin{equation}
\label{5.23}\vc{v}^+_{lmM}(\vc{r})=g_l(\alpha r)\vc{C}_{lm}(\hat{\vc{r}}),
\end{equation}
where $g_l(z)$ is the regular modified spherical Bessel function. The corresponding outgoing waves are defined by
\begin{equation}
\label{5.24}\vc{v}^-_{lmM}(\vc{r})=\frac{2\alpha}{\pi l(l+1)}k_l(\alpha r)\vc{C}_{lm}(\hat{\vc{r}}).
\end{equation}
The function $\vc{v}_{\Omega A\omega,0}(\vc{r})$ in Eq. (4.18) is proportional to $\vc{v}^-_{10M}(\vc{r})$, and
in analogy to Eq. (5.3) there is an addition theorem of the form
 \begin{equation}
\label{5.25}
\vc{v}^-_{l0M}(\vc{r})=\sum_{l'}S^{+-}(R;l'0M,l0M)\vc{v}^+_{l'0M}(\vc{r}_<).
\end{equation}
For the coefficients of low order in Eq. (5.22) one finds straightforwardly in the same way as before
 \begin{eqnarray}
\label{5.26}c^+_{10M}&=&2\sqrt{3\pi}\;\Omega_{A,0}\frac{a^3}{\alpha R^3}\frac{1+\alpha R}{1+\alpha a}\;e^{\alpha(a-R)},\nonumber\\
c^+_{20M}&=&-6\sqrt{5\pi}\;\Omega_{A,0}\frac{a^3}{\alpha^2R^4}\frac{B_0(\alpha R)}{1+\alpha a}\;e^{\alpha(a-R)}.
\end{eqnarray}
The first term in Eq. (5.22) yields the scalar mobility function $\alpha^{rr}_{BA,1}(R,\omega)$ given by Eq. (4.21). This corresponds to the first order angular velocity $\Omega_{B,1}$ of sphere $B$, given by
\begin{equation}
\label{5.27}\Omega_{B,1}=\frac{a^3}{R^3}\frac{1+\alpha R}{1+\alpha a}\;\frac{e^{\alpha(a+b-R)}}{B_0(\alpha b)}\;\Omega_{A,0}.
\end{equation}
The flow generated by sphere $B$ is given by
\begin{eqnarray}
\label{5.28}\vc{v}^-_{B1}(\vc{r})&=&\big(c^+_{10M}-\sqrt{\frac{4\pi}{3}}\frac{b}{g_1(\alpha b)}\Omega_{B,1}\big)\chi_{1MM}\vc{v}^-_{10M}(\vc{r}-\vc{R}_B)\nonumber\\
&+&\chi_{2MM}c^+_{20M}\vc{v}^-_{20M}(\vc{r}-\vc{R}_B),
\end{eqnarray}
with scattering coefficients
\begin{equation}
\label{5.29}
\chi_{lMM}=-\frac{\pi l (l+1)}{2\alpha}\frac{g_l(\alpha b)}{k_l(\alpha b)}.
\end{equation}
In the first line of Eq. (5.28) we took account of the fact \cite{12} that sphere $B$ is rotating with angular velocity $\Omega_{B,1}$.

The scattered wave acts back on sphere $A$ and the corresponding rotational velocity of sphere $A$, as its moves in this flow with zero torque, can be evaluated with the aid of Fax\'en's theorem Eq. (4.3) as above. As in Eq. (5.16)
\begin{equation}
\label{5.30}
\alpha^{rr}_{AA}(R,\omega)=\frac{1+\alpha a}{8\pi\eta a^3B_0(\alpha a)}+\alpha^{rr}_{AA,1}(R,\omega)+\alpha^{rr}_{AA,2}(R,\omega).
\end{equation}
The function $\alpha^{rr}_{AA,1}(R,\omega)$ is given by
\begin{equation}
\label{5.31}
\alpha^{rr}_{AA,1}(R,\omega)=\frac{P(\alpha b)}{16\pi\eta\alpha^3R^6B_0(\alpha a)^2B_0(\alpha b)}\;(1+\alpha R)^2e^{2\alpha(a-R)},
\end{equation}
with abbreviation $P(\lambda)$ given in the Appendix.
At low frequency the function behaves as
\begin{equation}
\label{5.32}
\alpha^{rr}_{AA,1}(R,\omega)=-\frac{b^5}{120\pi\eta R^6}\;\alpha^2+O(\alpha^4).
\end{equation}
This vanishes at zero frequency.
The function $\alpha^{rr}_{AA,2}(R,\omega)$ is given by
\begin{equation}
\label{5.33}
\alpha^{rr}_{AA,2}(R,\omega)=
\frac{45}{16\pi\eta\alpha^5R^8}\;\frac{P(\alpha b)}{B_0(\alpha a)^2B_0(\alpha b)}\;B_0(\alpha R)^2e^{2\alpha(a-R)},
\end{equation}
At low frequency this has the expansion
\begin{eqnarray}
\label{5.34}\alpha^{rr}_{AA,2}(R,\omega)=-\frac{3b^5}{8\pi\eta R^8}+O(\alpha^2).
\end{eqnarray}
The zero frequency limit agrees to terms of order $R^{-9}$ with earlier results \cite{4},\cite{5}.

\section{\label{VI}Transverse self-mobility functions}

The calculation of the transverse self-mobility functions is more complicated because of translation-rotation coupling, but in principle we can follow the same scheme as in Sec. V. In this section we derive approximations to the self-mobility functions $\beta^{tt}_{AA}(R,\omega)$ and $\beta^{rt}_{AA}(R,\omega)$. We consider the situation where to zero order sphere $A$ moves with velocity $U_{A,0}$ in the $x$ direction.

The zero order flow $\vc{v}_{UA\omega,0}(\vc{r})$ is given by Eqs. (4.4-6) with on the right hand side the subscript $z$ replaced by $x$. Correspondingly the functions in Eq. (4.8) are replaced by
\begin{eqnarray}
\label{6.1}\vc{u}_{1A}(\vc{r})&=&\sqrt{6\pi}\;\alpha^2a^3[\vc{v}^-_{1-1P}(\vc{r})-\vc{v}^-_{11P}(\vc{r})],\nonumber\\
\vc{v}_{1A}(\vc{r})&=&2\sqrt{6\pi}\;\frac{e^{\alpha a}}{\alpha}[\vc{v}^-_{1-1N}(\vc{r})-\vc{v}^-_{11N}(\vc{r})].
\end{eqnarray}
It is checked that the flow pattern $\vc{v}_{UA\omega,0}(\vc{r})$ satisfies the no-slip boundary condition
\begin{equation}
\label{6.2}\vc{v}_{UA\omega,0}(\vc{r})\big|_{r=a}=U_{A,0}\vc{e}_x.
\end{equation}
To first order sphere $B$ translates and rotates with zero force and torque in the flow pattern $\vc{v}_{UA\omega,0}(\vc{r})$. Sphere $B$ translates in the $x$ direction and rotates about the $y$ direction. From the Fax\'en theorems Eqs. (4.2) and (4.3) we find for the first order translational and rotational velocities of sphere $B$
\begin{eqnarray}
\label{6.3}U_{B,1}&=&-U_{A,0}\frac{3a}{2\alpha^2R^3A_0(\alpha b)}\bigg[B_0(\alpha a)B_0(\alpha b)-A_1(\alpha R)e^{\alpha(a+b-R)}\bigg],\nonumber\\
\Omega_{B,1}&=&-U_{A,0}\frac{3a}{4R^2B_0(\alpha b)}(1+\alpha R)e^{\alpha(a+b-R)},
\end{eqnarray}
The velocities $U_{B,1}$ and $\Omega_{B,1}$ were taken into account in Eqs. (4.14) and (4.16).

At $\vc{r}=\vc{R}_B$ we have
\begin{equation}
\label{6.4}\vc{v}_{UA\omega,0}(\vc{R}_B)=U_B^{(0)}\vc{e}_x,\qquad U_B^{(0)}=U_{A,0}\frac{3a}{2\alpha^2R^3}\big[-B_0(\alpha a)+A_1(\alpha R)e^{\alpha(a-R)}\big].
\end{equation}
Again we use an expansion of the form Eq. (5.4). The first term is
\begin{eqnarray}
\label{6.5}\vc{v}^{(0)}_{UAB}&=&c^+_{11N}\vc{v}^+_{11N}(\vc{r}-\vc{R}_B)+c^+_{1-1N}\vc{v}^+_{1-1N}(\vc{r}-\vc{R}_B)\nonumber\\
&+&c^+_{11P}\vc{v}^+_{11P}(\vc{r}-\vc{R}_B)+c^+_{1-1P}\vc{v}^+_{1-1P}(\vc{r}-\vc{R}_B),
\end{eqnarray}
with functions $\vc{v}^+_{lmN}(\vc{r})$ and $\vc{v}^+_{lmP}(\vc{r})$ defined in Eq. (5.2).

Comparing Eqs. (6.4) and (6.5) we find the relations
\begin{eqnarray}
\label{6.6}U^{(0)}_B&=&-c^+_{11N}\sqrt{\frac{2}{3\pi}}-c^+_{11P}\sqrt{\frac{3}{2\pi}},\nonumber\\
c^+_{1-1N}&=&-c^+_{11N},\qquad c^+_{1-1P}=-c^+_{11P}.
\end{eqnarray}
The Fax\'en theorem Eq. (4.2) must hold in the form
\begin{equation}
\label{6.7}B_0(\alpha b)U_B^{(0)}\vc{e}_x+B_2(\alpha b)b^2(\nabla^2\vc{v}^{(0)}_{UAB})\bigg|_{\vc{r}=\vc{R}_B}=A_0(\alpha b)U_{B,1}\vc{e}_x.
\end{equation}
Only the first line on the right in Eq. (6.5) contributes to the second term in this equation. Substituting we obtain the value of $c^+_{11N}$ as
\begin{equation}
\label{6.8}c^+_{11N}=-\frac{3}{2}\sqrt{\frac{3\pi}{2}}U_{A,0}\frac{a}{\alpha^2R^3}A_1(\alpha R)e^{\alpha(a-R)}.
\end{equation}
From Eq. (6.6) we find
\begin{equation}
\label{6.9}c^+_{11P}=\sqrt{\frac{3\pi}{2}}U_{A,0}\frac{a}{\alpha^2R^3}B_0(\alpha a).
\end{equation}
The corresponding scattered wave is
\begin{eqnarray}
\label{6.10}\vc{v}^-_{B1}(\vc{r})&=&c^+_{11N}\big[\chi_{1NN}\vc{v}^-_{11N}(\vc{r}-\vc{R}_B)+\chi_{1NP}\vc{v}^-_{11P}(\vc{r}-\vc{R}_B)\big]\nonumber\\
&+&\big(c^+_{11P}-\sqrt{\frac{4\pi}{3}}\;U_{B,11}\big)\big[\chi_{1NP}\vc{v}^-_{11N}(\vc{r}-\vc{R}_B)+\chi_{1PP}\vc{v}^-_{11P}(\vc{r}-\vc{R}_B)\big]\nonumber\\
&+&c^+_{1-1N}\big[\chi_{1NN}\vc{v}^-_{1-1N}(\vc{r}-\vc{R}_B)+\chi_{1NP}\vc{v}^-_{1-1P}(\vc{r}-\vc{R}_B)\big]\nonumber\\
&+&\big(c^+_{1-1P}-\sqrt{\frac{4\pi}{3}}\;U_{B,1-1}\big)\big[\chi_{1NP}\vc{v}^-_{1-1N}(\vc{r}-\vc{R}_B)+\chi_{1PP}\vc{v}^-_{10-1P}(\vc{r}-\vc{R}_B)\big],\nonumber\\
\end{eqnarray}
where $U_{B,1\pm 1}$ are spherical components of $U_{B,1}\vc{e}_x$. The scattered wave acts back on sphere $A$, and we can calculate its translational and rotational velocity by use of the Fax\'en theorems Eqs. (4.2) and (4.3). The translational velocity yields a contribution to the mobility function $\beta^{tt}_{AA}(R,\omega)$ given by
\begin{eqnarray}
\label{6.11}\beta^{tt}_{AA,1}(R,\omega)&=&\bigg[4\alpha^3b^3B_0(\alpha a)^2B_0(\alpha b)-8\alpha^3b^3B_0(\alpha a)A_1(\alpha R)e^{\alpha (a+b-R)}\nonumber\\
&-&(9-9\alpha b+\alpha^2b^2)A_1(\alpha R)^2e^{2\alpha(a+b-R)}+9A_0(\alpha b)A_1(\alpha R)^2e^{2\alpha(a-R)}\bigg]\nonumber\\
&\bigg/&\bigg[48\pi\eta\alpha^5R^6A_0(\alpha a)^2A_0(\alpha b)\bigg].
\end{eqnarray}
The function behaves at low frequency as
\begin{equation}
\label{6.12}\beta^{tt}_{AA,1}(R,\omega)=-\frac{b^5}{120\pi\eta R^6}+O(\alpha^2).
\end{equation}

The rotational velocity yields a contribution to the mobility
function $\beta^{rt}_{AA}(R,\omega)$ given by
\begin{eqnarray}
\label{6.13}\beta^{rt}_{AA,1}(R,\omega)&=&\frac{1+\alpha R}{96\pi\eta\alpha^3R^5A_0(\alpha
a)B_0(\alpha
a)A_0(\alpha b)}
\bigg[9\big(e^{2\alpha b}-1\big)A_0(\alpha b)A_1(\alpha R)e^{2\alpha(a-R)}\nonumber\\
&+&4\alpha^3b^3B_0(\alpha a)e^{\alpha(a+b-R)}-18\alpha bA_1(\alpha R)e^{2\alpha(a+b-R)}\bigg].
\end{eqnarray}
The function vanishes at low frequency as $\alpha^2$.

The next term in the expansion of the form Eq. (5.4) is
\begin{eqnarray}
\label{6.14}\vc{v}^{(1)}_{UAB}&=&c^+_{11M}\vc{v}^+_{11M}(\vc{r}-\vc{R}_B)+c^+_{1-1M}\vc{v}^+_{1-1M}(\vc{r}-\vc{R}_B)\nonumber\\
&+&c^+_{21P}\vc{v}^+_{21P}(\vc{r}-\vc{R}_B)+c^+_{2-1P}\vc{v}^+_{2-1P}(\vc{r}-\vc{R}_B).
\end{eqnarray}
The coefficients in Eq. (6.14) are found to be
\begin{eqnarray}
\label{6.15}c^+_{1\pm 1M}&=&-\frac{3i}{2}\sqrt{\frac{3\pi}{2}}\;U_{A,0}\frac{a(1+\alpha R)e^{\alpha(a-R)}}{\alpha R^2},\nonumber\\
c^+_{2\pm 1P}&=&\mp\frac{1}{2}\sqrt{\frac{3\pi}{10}}\;U_{A,0}\frac{a}{\alpha^2R^4}\big[6B_0(\alpha a)-B_3(\alpha R)e^{\alpha(a-R)}\big],
\end{eqnarray}
with abbreviation $B_3(\lambda)$ given in the Appendix.
From the theorem Eq. (4.2) we see that the flow $\vc{v}^{(1)}_{UAB}$ does not generate a translational velocity. From the theorem Eq. (4.3) we calculate the rotational velocity of sphere $B$, as it rotates in this flow with zero torque. This yields the mobility function given by Eq. (4.16).

The spherical components of the angular velocity of sphere $B$ are
\begin{equation}
\label{6.16} \Omega_{B,1\pm
1}=-\frac{3ia}{4\sqrt{2}}\frac{U_{A,0}}{R^2B_0(\alpha b)}(1+\alpha
R)e^{\alpha (a+b-R)},\qquad\Omega_{B,10}=0.
\end{equation}
The flow generated by sphere $B$ is
\begin{eqnarray}
\label{6.17}\vc{v}^-_{B2}(\vc{r})&=&\chi_{1MM}\bigg[\bigg(c^+_{11M}-\sqrt{\frac{4\pi}{3}}\frac{b}{g_1(\alpha b)}\Omega_{B,11}\bigg)\vc{v}^-_{11M}(\vc{r}-\vc{R}_B)\nonumber\\
&+&\bigg(c^+_{1-1M}-\sqrt{\frac{4\pi}{3}}\frac{b}{g_1(\alpha
b)}\Omega_{B,1-1}\bigg)\vc{v}^-_{1-1M}(\vc{r}-
\vc{R}_B)\bigg]\nonumber\\
&+&\chi_{2NP}\big[c^+_{21P}\vc{v}^-_{21N}(\vc{r}-\vc{R}_B)+c^+_{2-1P}\vc{v}^-_{2-1N}(\vc{r}-\vc{R}_B)\big]\nonumber\\
&+&\chi_{2PP}\big[c^+_{21P}\vc{v}^-_{21P}(\vc{r}-\vc{R}_B)+c^+_{2-1P}\vc{v}^-_{2-1P}(\vc{r}-\vc{R}_B)\big],
\end{eqnarray}
with outgoing waves given by Eqs. (4.9) and (5.24).

The wave in Eq. (6.17) acts back on sphere $A$ and we can
calculate its resulting translational and rotational velocity by
use of the theorems in Eqs. (4.2) and (4.3). From the
translational velocity we get a contribution to the mobility
function $\beta^{tt}_{AA}(R,\omega)$ given by
\begin{eqnarray}
\label{6.18}\beta^{tt}_{AA,2}(R,\omega)&=&\frac{-P(\alpha b)}{16\pi\eta\alpha^3R^4A_0(\alpha a)^2B_0(\alpha b)}(1+\alpha R)^2e^{2\alpha(a-R)}\nonumber\\
&-&\frac{b^3}{24\pi\eta\alpha^4R^8(1+\alpha b)A_0(\alpha a)^2}
\bigg[6B_0(\alpha a)-B_3(\alpha R)e^{\alpha(a-R)}\bigg]\nonumber\\
&\times&\bigg[2B_0(\alpha a)A_3(\alpha b)
-5B_3(\alpha R)e^{\alpha(a+b-R)}\bigg],
\end{eqnarray}
The first term derives from the the first two lines in Eq. (6.17).
At low frequency the function behaves as
\begin{equation}
\label{6.19}
\beta^{tt}_{AA,2}(R,\omega)=-\frac{5a^4b^3+3a^2b^5}{24\pi\eta
R^8}+O(\alpha^2).
\end{equation}
Here the first term in Eq, (6.18) does not contribute.
From the rotational velocity we get a contribution to the mobility function $\beta^{rt}_{AA}(R,\omega)$ given by
\begin{eqnarray}
\label{6.20}\beta^{rt}_{AA,2}(R,\omega)&=&\frac{-5b^3}{16\pi\eta\alpha^2R^7(1+\alpha b)}\frac{B_0(\alpha R)e^{\alpha(a+b-R)}}{A_0(\alpha a)B_0(\alpha a)}\bigg[6B_0(\alpha a)
-B_3(\alpha R)e^{\alpha(a-R)}\bigg].\nonumber\\
\end{eqnarray}
At low frequency this behaves as
\begin{equation}
\label{6.21}
\beta^{rt}_{AA,2}(R,\omega)=\frac{5a^2b^3}{16\pi\eta
R^7}+O(\alpha^2).
\end{equation}

We must also consider the second order term $\vc{v}^{(2)}_{UAB}$. This turns out to be given by
\begin{eqnarray}
\label{6.22}\vc{v}^{(2)}_{UAB}&=&
c^+_{21M}\vc{v}^+_{21M}(\vc{r}-\vc{R}_B)+c^+_{2-1M}\vc{v}^+_{2-1M}(\vc{r}-\vc{R}_B)\nonumber\\
&+&c^+_{31P}\vc{v}^+_{31P}(\vc{r}-\vc{R}_B)+c^+_{3-1P}\vc{v}^+_{3-1P}(\vc{r}-\vc{R}_B),
\end{eqnarray}
with coefficients
\begin{eqnarray}
\label{6.23}
c^+_{2\pm1M}&=&\frac{3ia}{2\alpha^2R^3}\sqrt{\frac{15\pi}{2}}\;U_{A0}B_0(\alpha R)e^{\alpha(a-R)},\nonumber\\
c^+_{3\pm1P}&=&\pm\frac{9a}{\alpha^2R^5}\sqrt{\frac{\pi}{21}}\;U_{A0}\bigg[15B_0(\alpha a)-B_4(\alpha R)e^{\alpha(a-R)}\big)\bigg],
\end{eqnarray}
with abbreviation $B_4(\lambda)$ given in the Appendix.

The wave generated by sphere $B$ is given by
\begin{eqnarray}
\label{6.24}\vc{v}^-_{B3}(\vc{r})
&=&c^+_{21M}\chi_{2MM}\vc{v}^-_{21M}(\vc{r}-\vc{R}_B)+c^+_{2-1M}\chi_{2MM}\vc{v}^-_{2-1M}(\vc{r}-\vc{R}_B)\nonumber\\
&+&c^+_{31P}\big[\chi_{3NP}\vc{v}^-_{31N}(\vc{r}-\vc{R}_B)+\chi_{3PP}\vc{v}^-_{31P}(\vc{r}-\vc{R}_B)]\nonumber\\
&+&c^+_{3-1P}\big[\chi_{3NP}\vc{v}^-_{3-1N}(\vc{r}-\vc{R}_B)+\chi_{3PP}\vc{v}^-_{3-1P}(\vc{r}-\vc{R}_B)].
\end{eqnarray}
From this wave acting on sphere $A$ located at the origin we can
evaluate the resulting translational velocity $U_{A,3}$ and
rotational velocity $\Omega_{A,3}$ by use of the Fax\'en theorems
in Eqs. (4.2) and (4.3). This yields a contribution to the
mobility function
\begin{eqnarray}
\label{6.25}
\beta^{tt}_{AA,3}(R,\omega)&=&\bigg[\alpha b^5\big(6B_0(\alpha a)A_4(\alpha b)\nonumber\\
&-&14e^{\alpha(a+b-R)}B_4(\alpha R)\big)\big(-45B_0(\alpha a)+B_4(\alpha R)e^{\alpha(a-R)}\big)\nonumber\\
&+&675P(\alpha b)e^{2\alpha(a-R)}R^4B_0(\alpha R)^2\bigg]
\bigg/\big[720\pi\eta\alpha^5R^{10}A_0(\alpha a)^2B_0(\alpha b)\big],
\end{eqnarray}
with abbreviation $A_4(\lambda)$ given in the Appendix.
At low frequency the function has the expansion
\begin{equation}
\label{6.26}\beta^{tt}_{AA,3}(R,\omega)=\frac{-175a^4b^5-75a^2b^7+70a^2b^5R^2+15b^7R^2-27b^5R^4}{160\pi\eta R^{10}}
+O(\alpha^2).
\end{equation}
The zero frequency limit, in combination with Eqs. (6.12) and (6.19), agrees to terms of order $R^{-7}$ with earlier results \cite{3},\cite{4},\cite{5}.

In our approximation the mobility function $\beta^{tt}_{AA}(R,\omega)$ is given by
\begin{equation}
\label{6.27}\beta^{tt}_{AA}(R,\omega)=\frac{1}{\zeta^t_A(\omega)}+\beta^{tt}_{AA,1}(R,\omega)+\beta^{tt}_{AA,2}(R,\omega)+\beta^{tt}_{AA,3}(R,\omega),
\end{equation}
with the last three contributions in Eqs. (6.11), (6.18), and (6.25).
The term linear in $\alpha$ in the expansion of this quantity in powers of frequency agrees with the exact relation Eq. (2.12), and arises only from the first term. The interaction terms decrease with distance as $1/R^6$ in the zero frequency limit.

The rotational velocity $\Omega_{A,3}$ yields a contribution to the mobility function $\beta^{rt}_{AA}$,
\begin{eqnarray}
\label{6.28}
\beta^{rt}_{AA,3}(R,\omega)&=&\bigg[225R^2P(\alpha b)B_0(\alpha R)B_3(\alpha R)e^{2\alpha(a-R)}\nonumber\\
&+&14\alpha^3b^5A_3(\alpha R)\big(45B_0(\alpha a)-e^{\alpha (a-R)}B_4(\alpha R)\big)e^{\alpha(a+b-R)}\bigg]\nonumber\\
&\bigg/&\big[1440\pi\eta\alpha^5R^9A_0(\alpha a)B_0(\alpha a)B_0(\alpha b)\big].
\end{eqnarray}
At low frequency this has the expansion
\begin{equation}
\label{6.29}\beta^{rt}_{AA,3}(R,\omega)=\frac{-35a^2b^5+3b^5R^2}{32\pi\eta R^{9}}+O(\alpha^2).
\end{equation}
The zero frequency limit, in combination with Eq. (6.21), agrees to terms of order $R^{-8}$ with earlier results \cite{4},\cite{5}. In our approximation the mobility function $\beta^{rt}_{AA}(R,\omega)$ is given by
\begin{equation}
\label{6.30}\beta^{rt}_{AA}(R,\omega)=\beta^{rt}_{AA,1}(R,\omega)+\beta^{rt}_{AA,2}(R,\omega)+\beta^{rt}_{AA,3}(R,\omega),
\end{equation}
with the three contributions in Eqs. (6.13), (6.20), and (6.28).

\section{\label{VII}Transverse rotational self-mobility functions}

In this section we derive approximations to the self-mobility functions $\beta^{tr}_{AA}(R,\omega)$ and $\beta^{rr}_{AA}(R,\omega)$. In the primary flow sphere $A$ rotates about the $x$ axis, with flow pattern given by Eq. (4.18) with $\vc{C}_{1z}$ replaced by $\vc{C}_{1x}=\vc{e}_x\times\vc{e}_r$. The zero order flow $\vc{v}_{\Omega A\omega,0}(\vc{r})$ is given by Eqs. (4.18-20) with subscript $z$ replaced by $x$. We expand this into multipole flows by use of a Taylor expansion about the point $\vc{R}_B$ in a form similar to Eq. (5.21),
\begin{equation}
\label{7.1}\vc{v}_{\Omega A\omega,0}(\vc{r})=\sum^\infty_{n=0}\vc{v}^{(n)}_{\Omega AB},
\end{equation}
where $n$ indicates the power of $\vc{r}-\vc{R}_B$ in the Taylor expansíon. The first term is
\begin{eqnarray}
\label{7.2}\vc{v}^{(0)}_{\Omega AB}&=&c^+_{11N}\vc{v}^+_{11N}(\vc{r}-\vc{R}_B)+c^+_{1-1N}\vc{v}^+_{1-1N}(\vc{r}-\vc{R}_B)\nonumber\\
&+&c^+_{11P}\vc{v}^+_{11P}(\vc{r}-\vc{R}_B)+c^+_{1-1P}\vc{v}^+_{1-1P}(\vc{r}-\vc{R}_B).
\end{eqnarray}
At $\vc{r}=\vc{R}_B$ we have
\begin{equation}
\label{7.3}\vc{v}_{\Omega A\omega,0}(\vc{R}_B)=U^{(0)}_B\vc{e}_y,\qquad U^{(0)}_B=-\frac{a^3}{R^2}\;\Omega_{A,0}\frac{1+\alpha R}{1+\alpha a}\;e^{\alpha(a-R)}.
\end{equation}
Comparing this with Eq. (7.2) we find the relations
\begin{eqnarray}
\label{7.4}U^{(0)}_B&=&-i\bigg[c^+_{11N}\sqrt{\frac{2}{3\pi}}+c^+_{11P}\sqrt{\frac{3}{2\pi}}\bigg],\nonumber\\
c^+_{1-1N}&=&c^+_{11N},\qquad c^+_{1-1P}=c^+_{11P}.
\end{eqnarray}
Sphere $B$ acquires a first order velocity $U_{B,1}$ in the $y$ direction with value
\begin{equation}
\label{7.5}U_{B,1}=-\frac{a^3}{R^2}\Omega_{A,0}\frac{1+\alpha R}{(1+\alpha a)A_0(\alpha b)}\;e^{\alpha(a+b-R)},
\end{equation}
as accounted for in Eq. (4.25).

The Fax\'en theorem Eq. (4.2) can be applied in the form
\begin{equation}
\label{7.6}B_0(\alpha b)U^{(0)}_B\vc{e}_y+B_2(\alpha b)b^2(\nabla^2\vc{v}^{(0)}_{\Omega AB})\bigg|_{\vc{r}=\vc{R}_B}=A_0(\alpha b)U_{B,1}\vc{e}_y.
\end{equation}
Only the first line on the right in Eq. (7.2) contributes to the second term in this equation. Substituting we obtain the value of $c^+_{11N}$ as
\begin{equation}
\label{7.7}c^+_{11N}=2i\sqrt{\frac{2\pi}{3}}\;\Omega_{A,0}\frac{\alpha^3a^3R(1+\alpha R)}{(1+\alpha a)Q(\alpha R)},
\end{equation}
with abbreviation $Q(\lambda)$ given in the Appendix.
The coefficient $c^+_{11P}$ is found from Eqs. (7.3) and (7.4) as
\begin{equation}
\label{7.8}c^+_{11P}=\frac{-i}{3}\sqrt{\frac{2\pi}{3}}\;\Omega_{A,0}\;a^3e^{\alpha a}\frac{(1+\alpha R)\big[3Q(\alpha R)e^{-\alpha R}+4\alpha^3R^3\big]}{R^2(1+\alpha a)Q(\alpha R)}.
\end{equation}

The wave $\vc{v}^-_{B1}(\vc{r})$ scattered by sphere $B$ has formally the same expression as Eq. (6.10). The scattered wave acts back on sphere $A$ and we can calculate its translational and rotational velocity by use of the Fax\'en theorems Eqs. (4.2) and (4.3). The translational velocity yields a contribution to the mobility function $\beta^{tr}_{AA}(R,\omega)$ given by
\begin{eqnarray}
\label{7.9}\beta^{tr}_{AA,1}(R,\omega)&=&\frac{1+\alpha R}{16\pi\eta\alpha^2R^5A_0(\alpha a)B_0(\alpha a)A_0(\alpha b)Q(\alpha R)}\nonumber\\
&\times &\bigg[
3b\big[A_1(\alpha R)e^{\alpha(a+b-R)}-B_0(\alpha a)B_0(\alpha b)\big]Q(\alpha R)e^{\alpha(a+b-R)}\nonumber\\
&+&2\alpha^2R^3A_0(\alpha b)\big(e^{2\alpha b}-1\big)A_1(\alpha R)e^{\alpha(2a-R)}\nonumber\\
&-&bA_0(\alpha b)\big(3Q(\alpha R)e^{-\alpha R}+4\alpha^3R^3\big)A_1(\alpha R)e^{\alpha(2a+b-R)}\nonumber\\
&+&bA_0(\alpha b)B_0(\alpha a)e^{\alpha a}\bigg(3B_0(\alpha b)Q(\alpha R)e^{-\alpha R}+4B_0(\alpha b)\alpha^3R^3-4e^{\alpha b}\alpha^3R^3\bigg)\bigg].\nonumber\\
\end{eqnarray}
The expression vanishes at zero frequency. The rotational velocity yields a contribution to the mobility function $\beta^{rr}_{AA}(R,\omega)$ given by
\begin{eqnarray}
\label{7.10}\beta^{rr}_{AA,1}(R,\omega)&=&\frac{(1+\alpha R)^2e^{\alpha (a-R)}}{32\pi\eta R^4A_0(\alpha b)B_0(\alpha a)^2Q(\alpha R)}\nonumber\\
&\times &\bigg[A_0(\alpha b)\bigg(2\alpha^2R^3e^{2\alpha a}\big(1-e^{2\alpha b}\big)+3bQ(\alpha R)e^{\alpha(a+b-R)}+4\alpha^3bR^3e^{\alpha(a+b)}\bigg)\nonumber\\
&-&3bQ(\alpha R)e^{\alpha(a+2b-R)}\bigg].
\end{eqnarray}
The expression vanishes at zero frequency.

At zero frequency the mobility functions $\beta^{tr}_{AA}(R,0)$ and $\beta^{rr}_{AA}(R,0)$ do not vanish. Therefore we consider also the term $\vc{v}^{(1)}_{\Omega AB}$ in the expansion Eq. (7.1). This takes the form of Eq. (6.14) with new coefficients, which are found to be
\begin{eqnarray}
\label{7.11}c^+_{1\pm 1M}&=&\pm\sqrt{\frac{3\pi}{2}}\;\Omega_{A,0}\frac{a^3}{\alpha R^3}\frac{A_1(\alpha R)}{1+\alpha a},\nonumber\\
c^+_{2\pm 1P}&=&i\sqrt{\frac{3\pi}{10}}\;\Omega_{A,0}\frac{a^3}{R^3}\frac{B_0(\alpha R)}{1+\alpha a}.
\end{eqnarray}
The flow generated by sphere $B$ takes the form of Eq. (6.17) with these coefficients and with spherical components of the angular velocity of sphere $B$ given by
\begin{equation}
\label{7.12} \Omega_{B,1\pm
1}=\pm\frac{1}{2\sqrt{2}}\;\Omega_{A,0}\frac{a^3A_1(\alpha R)}{R^3(1+\alpha
a)B_0(\alpha b)}\;e^{\alpha (a+b-R)},\qquad\Omega_{B,10}=0.
\end{equation}
The scattered wave acts back on sphere $A$ and we can calculate its translational and rotational velocity by use of the Fax\'en theorems Eqs. (4.2) and (4.3). The translational velocity yields a contribution to the mobility function $\beta^{tr}_{AA}(R,\omega)$ given by
\begin{eqnarray}
\label{7.13}\beta^{tr}_{AA,2}(R,\omega)&=&\frac{a^3e^{\alpha(a- R)}}{32\pi\eta\alpha^3R^7A_0(\alpha a)B_0(\alpha a)(1+\alpha b)B_0(\alpha b)}\nonumber\\
&\times &\bigg[4\alpha b^3B_0(\alpha a)B_0(\alpha b)A_3(\alpha b)B_0(\alpha R)\nonumber\\
&-&10\alpha b^3B_0(\alpha b)B_0(\alpha R)B_3(\alpha R)e^{\alpha(a+b-R)}\nonumber\\
&-&R^2(1+\alpha b)(1+\alpha R)A_1(\alpha R)P(\alpha b)e^{\alpha(a-R)}\bigg].
\end{eqnarray}
At low frequency the function behaves as
\begin{equation}
\label{7.14} \beta^{tr}_{AA,2}(R,\omega)=-\frac{5a^2b^3+3 b^5}{16\pi\eta R^7}+O(\alpha^2).
\end{equation}
The rotational velocity yields a contribution to the mobility function $\beta^{rr}_{AA}(R,\omega)$ given by
\begin{eqnarray}
\label{7.15}\beta^{rr}_{AA,2}(R,\omega)&=&\frac{-1}{64\pi\eta\alpha^3R^6B_0(\alpha a)^2(1+\alpha b)B_0(\alpha b)}\bigg[30\alpha^3b^3B_0(\alpha b)B_0(\alpha R)^2e^{\alpha(2a+b-2R)}\nonumber\\
&+&(1+\alpha b)P(\alpha b)A_1(\alpha R)^2e^{2\alpha(a-R)}\bigg].
\end{eqnarray}
At low frequency the function behaves as
\begin{equation}
\label{7.16} \beta^{rr}_{AA,2}(R,\omega)=-\frac{15b^3}{32\pi\eta R^6}+O(\alpha^2).
\end{equation}

We must also consider the second order term $\vc{v}^{(2)}_{\Omega AB}$. This turns out to be given by the expression on the right hand side of Eq. (6.22)
with coefficients
\begin{eqnarray}
\label{7.17}
c^+_{2\pm1M}&=&\mp\sqrt{\frac{5\pi}{6}}\;\Omega_{A,0}\frac{a^3}{\alpha^2R^4(1+\alpha a)}\;B_3(\alpha R)e^{\alpha(a-R)},\nonumber\\
c^+_{3\pm1P}&=&\frac{-2i}{15}\sqrt{\frac{\pi}{21}}\;\Omega_{A,0}\frac{a^3}{R^4(1+\alpha a)}A_3(\alpha R)\;e^{\alpha(a-R)}.
\end{eqnarray}
The wave generated by sphere $B$ is given by the expression on the right hand side of Eq. (6.24) with the present values for the coefficients.

From this wave acting on sphere $A$ located at the origin we can
evaluate the resulting translational velocity $U_{A,3}$ and
rotational velocity $\Omega_{A.3}$ by use of the Fax\'en theorems
in Eqs. (4.2) and (4.3). The translational velocity $U_{A,3}$ yields a contribution to the $\beta^{tr}_{AA}$ mobility function
\begin{eqnarray}
\label{7.18}
\beta^{tr}_{AA,3}(R,\omega)&=&\bigg[225R^2P(\alpha b)B_0(\alpha R)B_3(\alpha R)e^{2\alpha(a-R)}\nonumber\\
&-&14\alpha^3b^5B_4(\alpha R)A_3(\alpha R)e^{\alpha(2a+b-2R)}+6\alpha^3b^5B_0(\alpha a)A_4(\alpha b)A_3(\alpha R)e^{\alpha(a-R)}\bigg]\nonumber\\
&\bigg/&\big[1440\pi\eta\alpha^5R^9A_0(\alpha a)B_0(\alpha a)B_0(\alpha b)\big].
\end{eqnarray}
At low frequency the function has the expansion
\begin{equation}
\label{7.19}\beta^{tr}_{AA,3}(R,\omega)=-\frac{35a^2b^5+15b^7-3b^5R^2}{32\pi\eta R^{9}}+O(\alpha^2).
\end{equation}
The zero frequency limit, in combination with Eq. (7.14), agrees to terms of order $R^{-8}$ with earlier results \cite{4},\cite{5}. Note that the sum of Eqs. (6.21) and (6.29) is the negative of the sum of Eqs. (7.14) and (7.19) to terms of order $R^{-8}$.

A comparison of the expression for $\beta^{rt}_{AA}$ in Eq. (6.30) with the analogous result for $\beta^{tr}_{AA}$ which follows from Eqs. (7.9), (7.13), and (7.18) shows that in the approximation the symmetry relation $\beta^{rt}_{AA}=-\beta^{tr}_{AA}$ is not satisfied. However, it is satisfied in low powers of $\alpha$ and $R^{-1}$. This suggests that the expansion must be carried to sufficiently high order in $\vc{r}-\vc{R}_B$ to verify the symmetry.

The rotational velocity $\Omega_{A,3}$ yields a contribution to the $\beta^{rr}_{AA}$ mobility function
\begin{equation}
\label{7.20}
\beta^{rr}_{AA,3}(R,\omega)=\frac{e^{\alpha(2a+b-2R)}}{2880\pi\eta\alpha^5R^8B_0(\alpha a)^2B_0(\alpha b)}\bigg[75e^{-\alpha b}P(\alpha b)B_3(\alpha R)^2-14\alpha^5b^5A_3(\alpha R)^2
\bigg].
\end{equation}
At low frequency this has the expansion
\begin{equation}
\label{7.21}\beta^{rr}_{AA,3}(R,\omega)=-\frac{39b^5}{32\pi\eta R^8}+O(\alpha^2).
\end{equation}
The zero frequency limit, in combination with Eq. (7.16), agrees to terms of order $R^{-7}$ with earlier results \cite{4},\cite{5}.
The interaction terms decrease with distance as $1/R^6$ in the zero frequency limit.

\section{\label{VIII}Discussion of results}

In the above we derived detailed analytic results for the scalar mobility functions of two spheres immersed in a viscous incompressible fluid, based on the approximation of a single propagator between the spheres for the mutual functions, and a single reflection with two propagators for the self-functions. Finite-size corrections and sphere rotations were taken into account. It is clear that for large and intermediate distances the mutual functions dominate. By their nature the self-functions are analytically complicated. In practical application one will often restrict oneself to the relatively simple expressions for the mutual functions. The expressions for the self-functions will allow an estimate of the error one commits in doing this. The derivation presented in Sec. II shows how to find the motion of the spheres from the mobility functions. In the following we present some numerical estimates for the mobility functions.

First we compare the difference of the calculated translational mutual functions with the simple point approximation discussed in Sec. III. We consider two equal spheres of radius $a$ at center-to-center distance $R=5a$. As typical frequency we consider the value $\alpha a=1$. From Eqs. (3.3) and (4.12) we then find the ratio $\alpha^{tt}_{AB,1}/G_\parallel=1.203$. Surprisingly, at $R=10a$ the ratio is larger, and equal to $1.221$. Similarly we find from Eqs. (3.3) and (4.14) the ratio $\beta^{tt}_{AB,1}/G_\perp=1.106$. At $R=10a$ the ratio is $1.219$. These estimates show that for the mutual functions it is well worthwhile to take the finite-size corrections into account.

Next we consider the numerical relevance of the hydrodynamic interactions on the self-functions. Here we must compare the functions $\alpha^{tt}_{AA,1}$, and $\alpha^{tt}_{AA,2}$, given by Eqs. (5.17) and (5.19) with the single sphere value $1/\zeta^t_A$, given by Eq. (4.1). For $R=5a$ and $\alpha a=1$ we find
$\alpha^{tt}_{AA,1}\zeta^t_A=-0.00033$ and $\alpha^{tt}_{AA,2}\zeta^t_A=-0.00085$, showing that the correction terms are relatively small. Similarly we find from Eqs. (6.11), (6.18) and (6.25)
$\beta^{tt}_{AA,1}\zeta^t_A=0.00005$, $\beta^{tt}_{AA,2}\zeta^t_A=-0.00017$, and $\beta^{tt}_{AA,3}\zeta^t_A=-0.00003$. These values are even smaller.

We also compare the rotational self-functions $\alpha^{rr}_{AA,1}$ and $\alpha^{rr}_{AA,2}$, given by Eqs. (5.31) and (5.33) with the single sphere value $1/\zeta^r_A$, given by Eq. (4.1). For $R=5a$ and $\alpha a=1$ we find
$\alpha^{rr}_{AA,1}\zeta^r_A=-1.4\times 10^{-8}$ and $\alpha^{rr}_{AA,2}\zeta^r_A=-1.4\times 10^{-7}$, showing that the correction terms are small. Similarly we find from Eqs. (7.10), (7.15) and (7.20)
$\beta^{rr}_{AA,1}\zeta^r_A=-1.4\times 10^{-6}$, $\beta^{rr}_{AA,2}\zeta^r_A=-4.7\times 10^{-6}$, and $\beta^{rr}_{AA,3}\zeta^r_A=-1.5\times 10^{-6}$. Again these values are small.

It follows from Eqs. (4.16) and (4.25) that there are translation-rotation couplings of long range, provided the frequency is not too high. At higher frequency these couplings are exponentially screened.

We investigate the frequency-dependence of the $tt$ mutual mobility functions in some more detail. It follows from Eqs. (4.12) and (4.14) that at large distance the interaction is dipolar with a frequency-dependent amplitude factor given by the ratio $B_0(\alpha a)B_0(\alpha b)/( A_0(\alpha a)A_0(\alpha b))$. At zero frequency this takes the value unity, but at high frequency the factor tends to the value $9$, implying a significant dependence on frequency. The factor also characterizes the deviation from the simple Green function approximation Eq. (3.8).

We note that at zero frequency the translational mutual mobility functions are given by
\begin{eqnarray}
\label{8.1}
\alpha^{tt}_{AB,1}(R,0)&=&\frac{1}{4\pi\eta R}-\frac{a^2+b^2}{12\pi\eta R^3},\nonumber\\
\beta^{tt}_{AB,1}(R,0)&=&\frac{1}{8\pi\eta R}+\frac{a^2+b^2}{24\pi\eta R^3}.
\end{eqnarray}
This is just the Rotne-Prager hydrodynamic interaction \cite{40},\cite{41}, generalized to unequal spheres.  Thus the functions $\alpha^{tt}_{AB,1}(R,\omega)$ and $\beta^{tt}_{AB,1}(R,\omega)$ may be regarded as generalizations of the Rotne-Prager interaction to finite frequency. Similarly, Eqs. (4.21) and (4.23) are generalizations to non-zero frequency of the dipolar expression at zero frequency \cite{4},\cite{5}.

At fixed center-to-center distance $R$ all mobility functions depend strongly on frequency. This must be taken into account in applications. In particular, in the determination of the dynamic modulus of viscoelastic fluids via observed motion of a pair of particles \cite{33}, it is necessary to use an accurate approximation to the retarded hydrodynamic interactions.

Finally, we discuss briefly the effect of added mass. We restrict attention to motion along the axis of centers. It follows from Eqs. (2.7), (2.8), and (2.14) that in this case translational and rotational motions decouple. The friction coefficients for translational motion are given by
\begin{equation}
\label{8.2}\zeta^{tt\parallel}_{AA}(R,\omega)=\frac{\alpha^{tt}_{BB}}{\alpha^{tt}_{AA}\alpha^{tt}_{BB}-{\alpha^{tt}_{AB}}^2},
\qquad\zeta^{tt\parallel}_{AB}(R,\omega)=\frac{-\alpha^{tt}_{AB}}{\alpha^{tt}_{AA}\alpha^{tt}_{BB}-{\alpha^{tt}_{AB}}^2}.
\end{equation}
This shows that in the friction coefficients the mutual mobility function $\alpha^{tt}_{AB}(R,\omega)$ enters in the form of a geometric series. At large distance $R$ the mutual term $\alpha^{tt}_{AB}(R,\omega)$ is small in comparison with $\alpha^{tt}_{AA}(R,\omega)$ and $\alpha^{tt}_{BB}(R,\omega)$, which tend to $\mu^t_A=1/\zeta^t_A$ and $\mu^t_B=1/\zeta^t_B$, respectively. At high frequency the mutual term tends to zero asymptotically for large $R$ as $(a^3+b^3)/R^3$, at low frequency it tends to zero as $(a+b)/R$.

At fixed $R$ the denominators in Eq. (8.1) tend to zero at high frequency as $1/\alpha^4$, and the numerators tend to zero as $1/\alpha^2$, so that the friction coefficients increase in proportion to frequency, corresponding to an added mass in the equations of motion. The added mass depends on the distance $R$.
For transverse motions there is a similar added mass effect, but the situation is more complicated due to the translation-rotation coupling.

The added mass effect implies that at short time $t=0+$ the velocity correlation function $\vc{C}^{tt}_{AB}(t)$ of the two spheres does not vanish. The velocities are correlated, in contrast to the prediction of equipartition. The effect is similar to that studied by Zwanzig and Bixon \cite{42} for a single sphere. The effect was seen in the computer simulation of Tatsumi and Yamamoto \cite{32}. As discussed above, the use of the approximation Eq. (3.8) leads to prediction of the correlation effect smaller by a factor 9 than found by use of Eqs. (4.12) and (4.14).

\section{\label{IX}Mutual mobility functions for mixed slip-stick boundary conditions}

The derivation of Sec. IV can be extended without difficulty to the case where the fluid velocity satisfies mixed slip-stick boundary conditions on the surface of the two spheres. First we cast the known Fax\'en theorems for this case in the convenient form of Eqs. (4.2) and (4.3). Albano et al. \cite{43} derived the Fax\'en theorem for the force in terms of surface and volume averages of the incident flow. By use of the identities for the averages derived by Kim and Karrila \cite{8} the theorem can be cast in the convenient form
\begin{equation}
\label{9.1}\vc{K}_A=6\pi\eta a\big[B_0(\alpha a,\xi_A)\vc{v}_0+B_2(\alpha a,\xi_A)a^2\nabla^2\vc{v}_0\big]\big|_{\vc{r}=\vc{0}}-\zeta^t_A(\omega,\xi_A)\vc{U}_A,
\end{equation}
with slip parameter $\xi$, as defined earlier \cite{35}, and with functions as defined in the Appendix. The value $\xi=0$ corresponds to no-slip, and the value $\xi=1/3$ corresponds to perfect slip. The Fax\'en theorem for the torque which we derived earlier \cite{35},\cite{36} can be cast in the form
\begin{equation}
\label{9.2}\vc{T}_A=4\pi\eta a^3\frac{(1-3\xi_A)e^{\alpha a}}{1+\alpha a+\xi_A\alpha^2}(\nabla\times\vc{v}_0)\big|_{\vc{r}=\vc{0}}-\zeta^r_A(\omega,\xi_A)\vc{\Omega}_A.
\end{equation}

The primary flow of sphere $A$ moving with velocity $U_{A,0}$ in the $z$ direction is given by \cite{12}
\begin{equation}
\label{9.3}\vc{v}_{UA\omega,0}(\vc{r})=U_{A,0}\bigg[\frac{3B_0(\alpha a,\xi_A)}{2(1+\xi_A\alpha a)\alpha^2a^2}\;\vc{u}_{1A}(\vc{r})+\frac{1}{2}\;\alpha a\frac{1-\xi_A}{1+\xi_A\alpha a}\;\vc{v}_{1A}(\vc{r},\alpha)\bigg],
\end{equation}
and the primary flow of sphere $A$ rotating with angular velocity $\Omega_{A,0}$ about the $z$ direction is given by \cite{12}
\begin{equation}
\label{9.4}\vc{v}_{\Omega A\omega,0}(\vc{r})=\Omega_{A,0} a(1-3\xi_A)\frac{k_1(\alpha r)}{k_1(\alpha a)+\xi_Ae^{-\alpha a}}\vc{C}_{1z}(\hat{\vc{r}}).
\end{equation}

The derivation proceeds as in Sec. IV. We only quote the final expressions for the mutual mobility functions in one-propagator approximation. The longitudinal $tt$ mutual mobility function is given by
\begin{equation}
\label{9.5}\alpha^{tt}_{BA,1}(R,\omega)=\frac{(1+\xi_A\alpha a)(1+\xi_B\alpha b)B_0(\alpha a,\xi_A)B_0(\alpha b,\xi_B)-(1-\xi_A)(1-\xi_B)(1+\alpha R)e^{\alpha(a+b-R)}}{2\pi\eta\alpha^2R^3(1+\xi_A\alpha a)(1+\xi_B\alpha b)A_0(\alpha a,\xi_A)A_0(\alpha b,\xi_B)}.
\end{equation}
The transverse $tt$ mutual mobility function is given by
\begin{equation}
\label{9.6}\beta^{tt}_{BA,1}(R,\omega)=\frac{-(1+\xi_A\alpha a)(1+\xi_B\alpha b)B_0(\alpha a,\xi_A)B_0(\alpha b,\xi_B)+(1-\xi_A)(1-\xi_B)A_1(\alpha R)e^{\alpha(a+b-R)}}{4\pi\eta\alpha^2R^3(1+\xi_A\alpha a)(1+\xi_B\alpha b)A_0(\alpha a,\xi_A)A_0(\alpha b,\xi_B)}.
\end{equation}
These functions satisfy reciprocity. The term linear in $\alpha$ in an expansion in powers of $\alpha$ for both functions is given by $-\alpha/(6\pi\eta)$, as for no-slip. The longitudinal $rt$ mutual mobility function vanishes, and the transverse one is given by
\begin{equation}
\label{9.7}
\beta^{rt}_{BA}(R,\omega)=\frac{1-\xi_A}{1+\xi_A\alpha a}\frac{1}{8\pi\eta R^2A_0(\alpha a,\xi_A)B_0(\alpha b)}\;(1+\alpha R)e^{\alpha(a+b-R)}.
\end{equation}
The longitudinal $tr$ mutual mobility function vanishes, and the transverse one is found to be
\begin{equation}
\label{9.}
\beta^{tr}_{BA}(R,\omega)=-\frac{1-\xi_B}{1+\xi_B\alpha b}\frac{1}{8\pi\eta R^2B_0(\alpha a)A_0(\alpha b,\xi_B)}\;(1+\alpha R)e^{\alpha(a+b-R)}.
\end{equation}
in agreement with the reciprocity relation Eq. (2.15).

Both the longitudinal and transverse $rr$ mutual mobility functions are found to be independent of the slip coefficients $\xi_A,\xi_B$. They are therefore given by Eqs. (4.21) and (4.23). Earlier we found this remarkable property at zero frequency \cite{4},\cite{5}.

\section{\label{X}Conclusions}

The analytic expressions for retarded hydrodynamic interactions between two spheres will be useful in qualitative work in situations where the spheres are not very close. Existing schemes by which the interactions can be calculated accurately, also at short distances \cite{20}-\cite{22}, are cumbersome and not easily implemented. In particular the simple expressions for the mutual interactions, which we derived in Secs. IV and IX, will find practical use.

The retarded interactions are of interest in situations on the timescale of velocity relaxation, as well as in situations where oscillating forces or torques are applied. The theory of velocity relaxation and correlation functions can be applied in the discussion of computer simulations of Brownian motion \cite{32}. Oscillating forces occur in the numerical study of swimming of two unequal spheres \cite{44}-\cite{46}. Our theoretical analysis of this model took account of fluid inertia only as an added mass effect \cite{47}. It will be of interest to apply the present results to this model. The theory can also be applied to the three-sphere swimmer \cite{48}-\cite{51}, provided the hydrodynamic interaction can be approximated as a sum of pair interactions.

The mobility functions involving rotation are of interest in the application to ferrofluids \cite{52}. The long range of the hydrodynamic interaction suggests that it is relevant for magnetic relaxation.\\\\
$\vc{\mathrm{Acknowledgment}}$ I thank Prof. R. B. Jones for a critical reading of the manuscript.
\appendix

\newpage

\section{\label{A}}
In this Appendix we list the abbreviations used in the main text,
\begin{eqnarray}
\label{A.1}A_0(\lambda)&=&1+\lambda+\frac{1}{9}\lambda^2,\qquad
A_1(\lambda)=1+\lambda+\lambda^2,\nonumber\\
A_3(\lambda)&=&15+15\lambda+6\lambda^2+\lambda^3,\qquad
A_4(\lambda)=105+105\lambda+45\lambda^2+10\lambda^3+\lambda^4,\nonumber\\
B_0(\lambda)&=&1+\lambda+\frac{1}{3}\lambda^2,\qquad
B_2(\lambda)=\lambda^{-2}[e^\lambda-B_0(\lambda)],\nonumber\\
B_3(\lambda)&=&6+6\lambda+3\lambda^2+\lambda^3,\qquad
B_4(\lambda)=45+45\lambda+21\lambda^2+6\lambda^3+\lambda^4,\nonumber\\
P(\lambda)&=&3B_0(\lambda)-(3-3\lambda+\lambda^2)e^{2\lambda},\qquad
Q(\lambda)=A_1(\lambda)-(1-\lambda+\lambda^2)e^{2\lambda}.
\end{eqnarray}
These functions are relevant for no-slip boundary conditions. In Sec. IX we use related functions for mixed slip-stick boundary conditions with slip parameter $\xi$. These functions are
\begin{eqnarray}
\label{A.2}A_0(\lambda,\xi)&=&\frac{1}{1+\xi\lambda}\bigg[A_0(\lambda)-\xi\bigg(1+\lambda-\frac{1}{9}\lambda^3\bigg)\bigg],
\nonumber\\
B_0(\lambda,\xi)&=&\frac{1}{1+\xi\lambda}\bigg[B_0(\lambda)-\xi\bigg(1+\lambda-\frac{1}{3}\lambda^3\bigg)\bigg],
\nonumber\\
B_2(\lambda,\xi)&=&B_2(\lambda)+\xi\frac{(1+\lambda)(1+\lambda-e^\lambda)}{\lambda^2(1+\xi\lambda)}.
\end{eqnarray}
The no-slip condition corresponds to slip parameter $\xi=0$, and perfect slip corresponds to the value $\xi=1/3$. The friction coefficients of sphere $A$ are
\begin{eqnarray}
\label{A.3}\zeta^t_A(\omega,\xi_A)&=&6\pi\eta aA_0(\alpha a,\xi_A)=6\pi\eta a\bigg[(1-\xi_A)\frac{1+\alpha a}{1+\xi_A\alpha a}+\frac{1}{9}\alpha^2a^2\bigg],\nonumber\\
\zeta^r_A(\omega,\xi_A)&=&8\pi\eta a^3(1-3\xi_A)\;\frac{B_0(\alpha a)}{1+\alpha a+\xi_A\alpha^2a^2},
\end{eqnarray}
and similarly for sphere $B$.

\newpage

\end{document}